\documentclass[twocolumn,letterpaper,aps,prd,superscriptaddress,showpacs,floatfix]{revtex4}

\usepackage{graphicx}	
\usepackage{natbib}
\usepackage{multirow}
\usepackage{xspace}	

\newcommand{\pt}{\mbox{$p_T$}\xspace}

\newcommand{\sqs}{\mbox{$\sqrt{s}$}\xspace}

\newcommand{\pp}{\mbox{$p$$+$$p$}\xspace}

\begin{document}

\title{Low-mass vector-meson production at forward rapidity 
in $p$$+$$p$ collisions \\ at $\sqrt{s}=200$ GeV}

\newcommand{\abilene}{Abilene Christian University, Abilene, Texas 79699, USA}
\newcommand{\augie}{Department of Physics, Augustana College, Sioux Falls, South Dakota 57197, USA}
\newcommand{\banaras}{Department of Physics, Banaras Hindu University, Varanasi 221005, India}
\newcommand{\barc}{Bhabha Atomic Research Centre, Bombay 400 085, India}
\newcommand{\baruch}{Baruch College, City University of New York, New York, New York, 10010 USA}
\newcommand{\bnlcoll}{Collider-Accelerator Department, Brookhaven National Laboratory, Upton, New York 11973-5000, USA}
\newcommand{\bnlphys}{Physics Department, Brookhaven National Laboratory, Upton, New York 11973-5000, USA}
\newcommand{\caucr}{University of California - Riverside, Riverside, California 92521, USA}
\newcommand{\charlesczech}{Charles University, Ovocn\'{y} trh 5, Praha 1, 116 36, Prague, Czech Republic}
\newcommand{\chonbuk}{Chonbuk National University, Jeonju, 561-756, Korea}
\newcommand{\ciae}{Science and Technology on Nuclear Data Laboratory, China Institute of Atomic Energy, Beijing 102413, P.~R.~China}
\newcommand{\cns}{Center for Nuclear Study, Graduate School of Science, University of Tokyo, 7-3-1 Hongo, Bunkyo, Tokyo 113-0033, Japan}
\newcommand{\colorado}{University of Colorado, Boulder, Colorado 80309, USA}
\newcommand{\columbia}{Columbia University, New York, New York 10027 and Nevis Laboratories, Irvington, New York 10533, USA}
\newcommand{\czechtech}{Czech Technical University, Zikova 4, 166 36 Prague 6, Czech Republic}
\newcommand{\dapnia}{Dapnia, CEA Saclay, F-91191, Gif-sur-Yvette, France}
\newcommand{\elte}{ELTE, E{\"o}tv{\"o}s Lor{\'a}nd University, H - 1117 Budapest, P{\'a}zm{\'a}ny P. s. 1/A, Hungary}
\newcommand{\ewha}{Ewha Womans University, Seoul 120-750, Korea}
\newcommand{\fsu}{Florida State University, Tallahassee, Florida 32306, USA}
\newcommand{\gsu}{Georgia State University, Atlanta, Georgia 30303, USA}
\newcommand{\hanyang}{Hanyang University, Seoul 133-792, Korea}
\newcommand{\hiroshima}{Hiroshima University, Kagamiyama, Higashi-Hiroshima 739-8526, Japan}
\newcommand{\howard}{Department of Physics and Astronomy, Howard University, Washington, DC 20059, USA}
\newcommand{\ihepprot}{IHEP Protvino, State Research Center of Russian Federation, Institute for High Energy Physics, Protvino, 142281, Russia}
\newcommand{\illuiuc}{University of Illinois at Urbana-Champaign, Urbana, Illinois 61801, USA}
\newcommand{\inrras}{Institute for Nuclear Research of the Russian Academy of Sciences, prospekt 60-letiya Oktyabrya 7a, Moscow 117312, Russia}
\newcommand{\instpasczech}{Institute of Physics, Academy of Sciences of the Czech Republic, Na Slovance 2, 182 21 Prague 8, Czech Republic}
\newcommand{\isu}{Iowa State University, Ames, Iowa 50011, USA}
\newcommand{\jaea}{Advanced Science Research Center, Japan Atomic Energy Agency, 2-4 Shirakata Shirane, Tokai-mura, Naka-gun, Ibaraki-ken 319-1195, Japan}
\newcommand{\jyvaskyla}{Helsinki Institute of Physics and University of Jyv{\"a}skyl{\"a}, P.O.Box 35, FI-40014 Jyv{\"a}skyl{\"a}, Finland}
\newcommand{\kek}{KEK, High Energy Accelerator Research Organization, Tsukuba, Ibaraki 305-0801, Japan}
\newcommand{\korea}{Korea University, Seoul, 136-701, Korea}
\newcommand{\kurchatov}{Russian Research Center ``Kurchatov Institute", Moscow, 123098 Russia}
\newcommand{\kyoto}{Kyoto University, Kyoto 606-8502, Japan}
\newcommand{\labllr}{Laboratoire Leprince-Ringuet, Ecole Polytechnique, CNRS-IN2P3, Route de Saclay, F-91128, Palaiseau, France}
\newcommand{\lahorelums}{Physics Department, Lahore University of Management Sciences, Lahore 54792, Pakistan}
\newcommand{\lawllnl}{Lawrence Livermore National Laboratory, Livermore, California 94550, USA}
\newcommand{\losalamos}{Los Alamos National Laboratory, Los Alamos, New Mexico 87545, USA}
\newcommand{\lpc}{LPC, Universit{\'e} Blaise Pascal, CNRS-IN2P3, Clermont-Fd, 63177 Aubiere Cedex, France}
\newcommand{\lund}{Department of Physics, Lund University, Box 118, SE-221 00 Lund, Sweden}
\newcommand{\maryland}{University of Maryland, College Park, Maryland 20742, USA}
\newcommand{\mass}{Department of Physics, University of Massachusetts, Amherst, Massachusetts 01003-9337, USA }
\newcommand{\michigan}{Department of Physics, University of Michigan, Ann Arbor, Michigan 48109-1040, USA}
\newcommand{\muenster}{Institut fur Kernphysik, University of Muenster, D-48149 Muenster, Germany}
\newcommand{\muhlenberg}{Muhlenberg College, Allentown, Pennsylvania 18104-5586, USA}
\newcommand{\myongji}{Myongji University, Yongin, Kyonggido 449-728, Korea}
\newcommand{\nagasaki}{Nagasaki Institute of Applied Science, Nagasaki-shi, Nagasaki 851-0193, Japan}
\newcommand{\newmex}{University of New Mexico, Albuquerque, New Mexico 87131, USA }
\newcommand{\nmsu}{New Mexico State University, Las Cruces, New Mexico 88003, USA}
\newcommand{\ohio}{Department of Physics and Astronomy, Ohio University, Athens, Ohio 45701, USA}
\newcommand{\ornl}{Oak Ridge National Laboratory, Oak Ridge, Tennessee 37831, USA}
\newcommand{\orsay}{IPN-Orsay, Universite Paris Sud, CNRS-IN2P3, BP1, F-91406, Orsay, France}
\newcommand{\peking}{Peking University, Beijing 100871, P.~R.~China}
\newcommand{\pnpi}{PNPI, Petersburg Nuclear Physics Institute, Gatchina, Leningrad region, 188300, Russia}
\newcommand{\riken}{RIKEN Nishina Center for Accelerator-Based Science, Wako, Saitama 351-0198, Japan}
\newcommand{\rikjrbrc}{RIKEN BNL Research Center, Brookhaven National Laboratory, Upton, New York 11973-5000, USA}
\newcommand{\rikkyo}{Physics Department, Rikkyo University, 3-34-1 Nishi-Ikebukuro, Toshima, Tokyo 171-8501, Japan}
\newcommand{\saispbstu}{Saint Petersburg State Polytechnic University, St. Petersburg, 195251 Russia}
\newcommand{\saopaulo}{Universidade de S{\~a}o Paulo, Instituto de F\'{\i}sica, Caixa Postal 66318, S{\~a}o Paulo CEP05315-970, Brazil}
\newcommand{\seoulnat}{Department of Physics and Astronomy, Seoul National University, Seoul 151-742, Korea}
\newcommand{\stonybrkc}{Chemistry Department, Stony Brook University, SUNY, Stony Brook, New York 11794-3400, USA}
\newcommand{\stonycrkp}{Department of Physics and Astronomy, Stony Brook University, SUNY, Stony Brook, New York 11794-3800, USA}
\newcommand{\tenn}{University of Tennessee, Knoxville, Tennessee 37996, USA}
\newcommand{\titech}{Department of Physics, Tokyo Institute of Technology, Oh-okayama, Meguro, Tokyo 152-8551, Japan}
\newcommand{\tsukuba}{Institute of Physics, University of Tsukuba, Tsukuba, Ibaraki 305, Japan}
\newcommand{\vandy}{Vanderbilt University, Nashville, Tennessee 37235, USA}
\newcommand{\weizmann}{Weizmann Institute, Rehovot 76100, Israel}
\newcommand{\wigner}{Institute for Particle and Nuclear Physics, Wigner Research Centre for Physics, Hungarian Academy of Sciences (Wigner RCP, RMKI) H-1525 Budapest 114, POBox 49, Budapest, Hungary}
\newcommand{\yonsei}{Yonsei University, IPAP, Seoul 120-749, Korea}
\newcommand{\zagreb}{University of Zagreb, Faculty of Science, Department of Physics, Bijeni\v{c}ka 32, HR-10002 Zagreb, Croatia}
\affiliation{\abilene}
\affiliation{\augie}
\affiliation{\banaras}
\affiliation{\barc}
\affiliation{\baruch}
\affiliation{\bnlcoll}
\affiliation{\bnlphys}
\affiliation{\caucr}
\affiliation{\charlesczech}
\affiliation{\chonbuk}
\affiliation{\ciae}
\affiliation{\cns}
\affiliation{\colorado}
\affiliation{\columbia}
\affiliation{\czechtech}
\affiliation{\dapnia}
\affiliation{\elte}
\affiliation{\ewha}
\affiliation{\fsu}
\affiliation{\gsu}
\affiliation{\hanyang}
\affiliation{\hiroshima}
\affiliation{\howard}
\affiliation{\ihepprot}
\affiliation{\illuiuc}
\affiliation{\inrras}
\affiliation{\instpasczech}
\affiliation{\isu}
\affiliation{\jaea}
\affiliation{\jyvaskyla}
\affiliation{\kek}
\affiliation{\korea}
\affiliation{\kurchatov}
\affiliation{\kyoto}
\affiliation{\labllr}
\affiliation{\lahorelums}
\affiliation{\lawllnl}
\affiliation{\losalamos}
\affiliation{\lpc}
\affiliation{\lund}
\affiliation{\maryland}
\affiliation{\mass}
\affiliation{\michigan}
\affiliation{\muenster}
\affiliation{\muhlenberg}
\affiliation{\myongji}
\affiliation{\nagasaki}
\affiliation{\newmex}
\affiliation{\nmsu}
\affiliation{\ohio}
\affiliation{\ornl}
\affiliation{\orsay}
\affiliation{\peking}
\affiliation{\pnpi}
\affiliation{\riken}
\affiliation{\rikjrbrc}
\affiliation{\rikkyo}
\affiliation{\saispbstu}
\affiliation{\saopaulo}
\affiliation{\seoulnat}
\affiliation{\stonybrkc}
\affiliation{\stonycrkp}
\affiliation{\tenn}
\affiliation{\titech}
\affiliation{\tsukuba}
\affiliation{\vandy}
\affiliation{\weizmann}
\affiliation{\wigner}
\affiliation{\yonsei}
\affiliation{\zagreb}
\author{A.~Adare} \affiliation{\colorado}
\author{C.~Aidala} \affiliation{\losalamos} \affiliation{\michigan}
\author{N.N.~Ajitanand} \affiliation{\stonybrkc}
\author{Y.~Akiba} \affiliation{\riken} \affiliation{\rikjrbrc}
\author{R.~Akimoto} \affiliation{\cns}
\author{H.~Al-Ta'ani} \affiliation{\nmsu}
\author{J.~Alexander} \affiliation{\stonybrkc}
\author{M.~Alfred} \affiliation{\howard}
\author{K.R.~Andrews} \affiliation{\abilene}
\author{A.~Angerami} \affiliation{\columbia}
% ...
\collaboration{PHENIX Collaboration} \noaffiliation

\date{\today}


\begin{abstract}


The PHENIX experiment at the Relativistic Heavy Ion Collider has 
measured low mass vector meson, $\omega$, $\rho$, and $\phi$, production 
through the dimuon decay channel at forward rapidity ($1.2<|y|<2.2$) 
in $p$$+$$p$ collisions at $\sqrt{s}=200$~GeV. The differential cross 
sections for these mesons are measured as a function of both $p_T$ and 
rapidity.  We also report the integrated differential cross sections over 
$1<p_T<7$ GeV/$c$ and $1.2<|y|<2.2$:  
$d\sigma/dy(\omega+\rho\rightarrow\mu\mu) =
80 \pm 6 \mbox{ (stat)} \pm 12 \mbox{ (syst)}$~nb
and
$d\sigma/dy(\phi\rightarrow\mu\mu) = 
27 \pm 3 \mbox{ (stat)} \pm 4 \mbox{ (syst)}$~nb.
These results are compared with midrapidity measurements and calculations.

\end{abstract}

\pacs{13.20.Jf, 25.75.Dw} 
	

\maketitle

\section{Introduction}

Low-mass vector meson (LVM) production in \pp collisions is an important 
tool to study quantum chromodynamics (QCD), providing data to tune 
phenomenological soft QCD models and to compare to hard perturbative QCD 
calculations. Various 
experiments~\cite{STAR_Phi2,PHENIX_pp_200_central,E735_ppbar_phi,ALICE_pp_Phi_central,ALICE_pp_forward,LHCb_pp_Phi} 
have studied LVM at different colliding energies and in different 
kinematic regions.

In addition, LVM production in \pp collisions provides a reference for 
high-energy heavy-ion-collision measurements. LVM studies provide key 
information on the hot and dense state of the strongly interacting matter 
produced in such collisions. Among them, strangeness 
enhancement~\cite{Koch}, a phenomenon associated with soft particles in 
bulk matter, can be accessed through the measurements of $\phi$-meson 
production~\cite{NA49_phi,NA50_Phi, CERES_Phi, NA60_Phi, 
PHENIX_Phi_RAA_200_central,STAR_Phi} and the $\phi$/($\rho+\omega$) ratio. 
The measurement of the $\rho$ spectral function can be used to reveal 
in-medium modifications of the hadron properties close to the QCD phase 
boundary linked to chiral symmetry 
restoration~\cite{vanHees,CERES,NA60_Rho}.  However, measuring the $\rho$ 
spectral function in the two-muon channel requires better mass resolution 
than is provided by the muon spectrometers of the PHENIX experiment at the 
Relativistic Heavy Ion Collider.

Having two muon spectrometers covering the rapidity range $1.2<|y|<2.2$, 
PHENIX is able to study vector-meson production via the dimuon decay 
channel.  Because there is no similar measurement in this kinematic regime 
at this energy, the forward rapidity measurements are a valuable 
addition to the database and are complementary to previously published 
midrapidity results~\cite{STAR_Phi2,PHENIX_pp_200_central}.  We report the 
differential cross section as a function of \pt and rapidity of 
$(\omega+\rho)$ and $\phi$ mesons for $1<\pt<7$~GeV/$c$ and $1.2<|y|<2.2$. 
Results presented in this paper are based on the data sample collected in 
2009 using the PHENIX muon spectrometers in \pp collisions at 
\sqs~=~200~GeV. The sampled luminosity of the data used in this analysis 
corresponds to 14.1 pb$^{-1}$.

\section{Experiment}

The PHENIX apparatus is described in detail in~\cite{NIMA.499.469}. This 
analysis uses the dimuon decay channel of the low mass vector mesons. The 
detectors relevant for reconstruction and triggering are the two muon 
spectrometers~\cite{NIMA.499.480} and the two beam-beam counters (BBCs) 
in the forward and backward rapidities.

The muon spectrometers, located behind an absorber composed of 19 cm 
copper and 60 cm iron, include the muon tracker (MuTr), which is in a 
radial magnetic field with an integrated bending power of 0.8 Tesla-meter, 
followed by the muon identifier (MuID). The muon spectrometers cover the 
range $1.2<|\eta|<2.2$ over the full azimuth. The MuTr comprises 
three sets of cathode strip chambers while the MuID comprises five 
planes of Iarocci tubes interleaved with steel absorber plates. The 
composite momentum resolution, $\delta p/p$, of particles in the analyzed 
momentum range is about 5\% independent of momentum and dominated by 
multiple scattering, and the LVM mass resolution is 85 MeV/$c^2$. Muon 
candidates are identified by reconstructed tracks in the MuTr matched to 
MuID tracks that penetrate through to the last MuID plane. The minimum 
momentum of a muon to reach the last MuID plane is $\sim$2 GeV/$c$.

Beam-beam counters (BBC), consisting of two arrays of 64 \v{C}erenkov 
counters covering the pseudorapidity range $3.1<|\eta|<3.9$, were used 
to measure the collision vertex along the beam axis ($z_\mathrm{vtx}$) 
with 2-cm resolution in addition to providing a minimum-bias trigger.

\section{Data Analysis}

The data set for this analysis was recorded in 2009 using a minimum-bias 
trigger that required at least one hit in each of the BBCs. Additionally, 
the MuID Level-1 dimuon trigger was used which required that at least two 
tracks penetrate through the MuID to its last layer.

A set of quality assurance cuts is applied to the data to select good muon 
candidates and improve the signal to background ratio. The BBC collision z 
vertex is required to be within $\pm$30 cm of the center of the 
interaction region along the beam direction. The MuTr tracks are matched 
to the MuID tracks at the first MuID layer in both position and angle. In 
addition, the track trajectory is required to have at least 8 of 10 
possible hits in the MuID.

The invariant mass distribution is formed by combining muon candidate 
tracks of opposite charge. In addition to low mass vector mesons, the 
invariant mass spectra contains uncorrelated and correlated backgrounds. 
The uncorrelated backgrounds arise from random combinatoric associations 
of unrelated muons candidates while the correlated backgrounds arise from 
open charm decay (e.g., $D\bar{D}$ where both decay semileptonically to 
muons), open bottom decay, $\eta$ and $\omega$ Dalitz decays and the 
Drell-Yan process.

\begin{figure}
\includegraphics[width=1.0\linewidth]{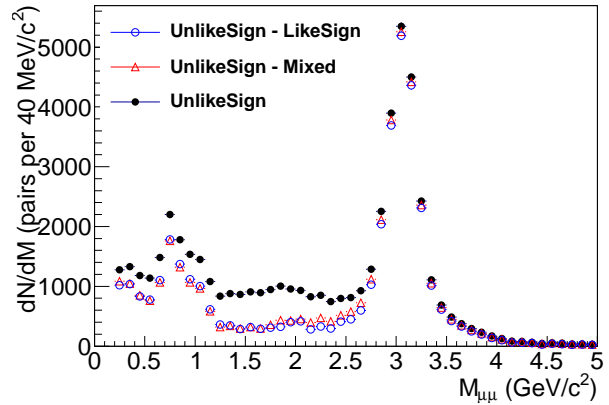}
\caption{\label{fig:FGBG} (color online) 
The unlike-sign dimuon invariant mass spectrum before background 
subtraction (solid black points), after subtracting mixed events 
background (empty red triangles) and after subtracting like-sign 
background (empty blue circles).
}
\end{figure}

Traditionally, the combinatorial background is estimated and subtracted by 
two methods. The first method uses the mass spectra of the like-sign pairs 
that are reconstructed within the same event. The other forms unlike-sign 
and like-sign pairs from different events and is often referred to as the 
``mixed-event method.'' In the like-sign method, the like-sign pairs are 
expected to originate from combinatorial processes; in addition there can 
be correlated pairs within a single event~\cite{PhysRevC.84.054912}. In 
the case of the mixed event method, unlike-sign pairs are formed from 
tracks from different events which provides purely combinatorial 
pairs~\cite{PhysRevC.84.054912,PhysRevC.87.034904}. The results of using 
these two methods are shown in Fig.~\ref{fig:FGBG}.

It is clear from Fig.~\ref{fig:FGBG} that the two methods are not able 
to reproduce the background in the low mass region. Hence, we introduce a 
new data driven technique here.

The background below 1.4 GeV/$c^{2}$ is dominated by 
\begin{enumerate}
\item $K/\pi\rightarrow\mu$ decays that occur before reaching the absorber 
\item punch-through hadrons with high \pt that are misidentified as muons 
and
\item muons that result from decay in the muon tracker volume. 
\end{enumerate}
A $\chi^{2}$ statistic is calculated from a simultaneous fit of the
two muon tracks with a common event determined by the BBC.  Tracks due
to the backgrounds listed above produce a broader  $\chi^{2}$ distribution
than that of true muon tracks, and this difference can be used to
discriminate statistically between foregrounds and backgrounds.  We
classify pairs with $\chi^{2}_\mathrm{vtx}<3.6$ as foreground pairs
and those with $\chi^{2}_\mathrm{vtx}>3.6$ as background pairs.
The value, $\chi^{2}_{\rm vtx,cut}=3.6$, was selected such that we 
retain as much of the signal as possible, while still allowing enough 
statistics in our background sample.

\begin{figure}
\includegraphics[width=1.0\linewidth]{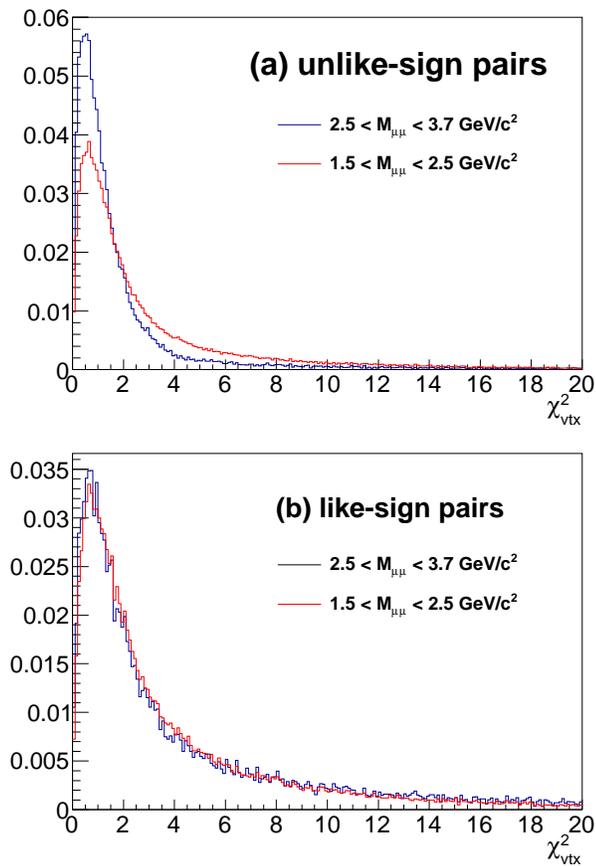}
\caption{\label{fig:chi2vtx} (color online) 
The $\chi^{2}_\mathrm{vtx}$ distributions for nonresonant mass region (red), 
and signal ($J/\psi$) mass region (blue) The unlike-sign pairs are shown 
in (a) while the like-sign pairs are shown in (b). In each panel, the 
histograms are normalized to the total number of events.
}
\end{figure}

\begin{figure}
\includegraphics[width=1.0\linewidth]{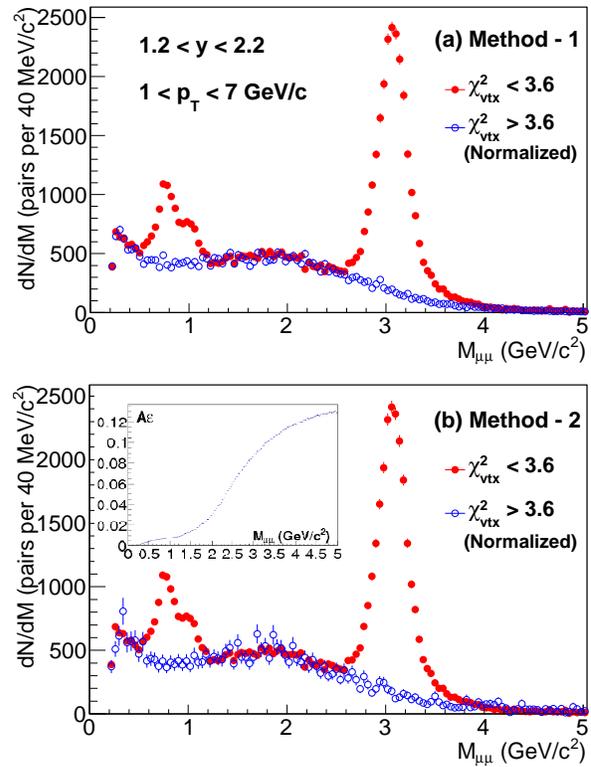}
\caption{\label{fig:bknorm} (color online) 
The unlike-sign dimuon invariant mass spectrum (solid red points) and the 
background spectrum (empty blue circles) normalized using the first 
normalization method in (a) while using the second normalization method in 
(b). The insert in (b) shows the muon arms acceptance and reconstruction 
efficiency.}
\end{figure}

Figure~\ref{fig:chi2vtx}(a) shows the unlike-sign pairs 
$\chi^{2}_\mathrm{vtx}$ distribution, which is narrower in the resonance 
region dominated by prompt dimuons (e.g, in the $J/\psi$ region, 
$2.5<M_{\mu^+\mu^-}<3.7$~GeV/$c^2$), and wider in the nonresonant regions. 
On the other hand, the $\chi^{2}_\mathrm{vtx}$ distribution for the 
like-sign pairs is the same in both mass regions. In addition, the 
unlike-sign $\chi^{2}_\mathrm{vtx}$ distribution matches very well that of 
the like-sign in the nonresonant region. After selecting the foreground 
and background from the data, the background is normalized to the 
foreground by two normalization methods: The first method uses the 
unlike-sign pairs, where the ratio of the foreground to background spectra 
is fitted by a polynomial in the nonresonant region and the background 
spectra are then multiplied by the fit function. The other uses the ratio 
of like-sign pairs corresponding to 
$\chi^{2}_\mathrm{vtx}<\chi^{2}_\mathrm{vtx,cut}$ and 
$\chi^{2}_\mathrm{vtx}>\chi^{2}_\mathrm{vtx,cut}$ to determine the 
functional form of the shape of the background. This function is then 
normalized to match the unlike-sign distribution in the nonresonant 
regions, $0.3<M_{\mu^+\mu^-}<0.6$~GeV/$c^2$ and 
$1.5<M_{\mu^+\mu^-}<2.5$~GeV/$c^2$. Background estimates using those two 
methods are shown in Fig.~\ref{fig:bknorm}.

Both estimates of the background match the nonresonant region of the 
unlike-sign spectrum. However, because the second method includes a two-step 
normalization which introduces higher statistical fluctuations upon the 
background subtraction, the second method is only used for a cross check. 
The insert in Fig.~\ref{fig:bknorm}(b) shows that the acceptance and 
reconstruction efficiency drops quickly at low mass which explains the 
higher $J/\psi$ yield compared to the low mass vector mesons.

To ensure the robustness of the yield extraction, an additional yield 
extraction procedure is employed. The background is fitted with a 
polynomial and the result of the fit is added to signal fits which are 
then fitted to the dimuon invariant mass spectrum while constraining the 
added function with the background spectra fit parameters. The background 
normalization is a free parameter.

The unlike-sign dimuon spectra, with 
$\chi^{2}_\mathrm{vtx}<\chi^{2}_\mathrm{vtx,cut}$, in the region of 
interest ($0<M_{\mu^+\mu^-}<2$ GeV/$c^2$) have contributions from 
three mesons, $\omega$, $\rho$, and $\phi$. The $\phi$ meson is partly 
resolved while $\omega$ and $\rho$ mesons are completely merged, hence the 
combined yield for $\omega$ and $\rho$ mesons was extracted. It was found 
that the reconstructed mass spectra of the simulated $\omega$ and $\phi$ 
are fitted well by Gaussian distributions, while in the case of $\rho$, a 
Breit Wigner distribution matched the mass spectrum, which motivated using 
these distributions to fit the invariant mass spectra.

The background subtracted dimuon spectra in the low mass region, 
$0.3<M_{\mu^+\mu^-}<2.5$ GeV/$c^2$, are fitted with two Gaussian 
distributions and a Breit Wigner distribution. The means and widths 
($\Gamma$ for Breit Wigner distribution) of the reconstructed $\omega$, 
$\rho$ and $\phi$ were extracted using the PHENIX simulation chain and 
used as a first approximation in fitting the data. The masses and widths 
are free parameters in the fit to account for small detector effects which 
result in $<2$\% variations with respect to the PDG values. In addition to 
these distributions, the dimuon spectra without background subtraction are 
fitted with a polynomial. It is important to note that the parameters from 
data and simulation fits converged to the same values within uncertainties 
without any systematic shifts.

\begin{figure*}
\includegraphics[width=0.998\linewidth]{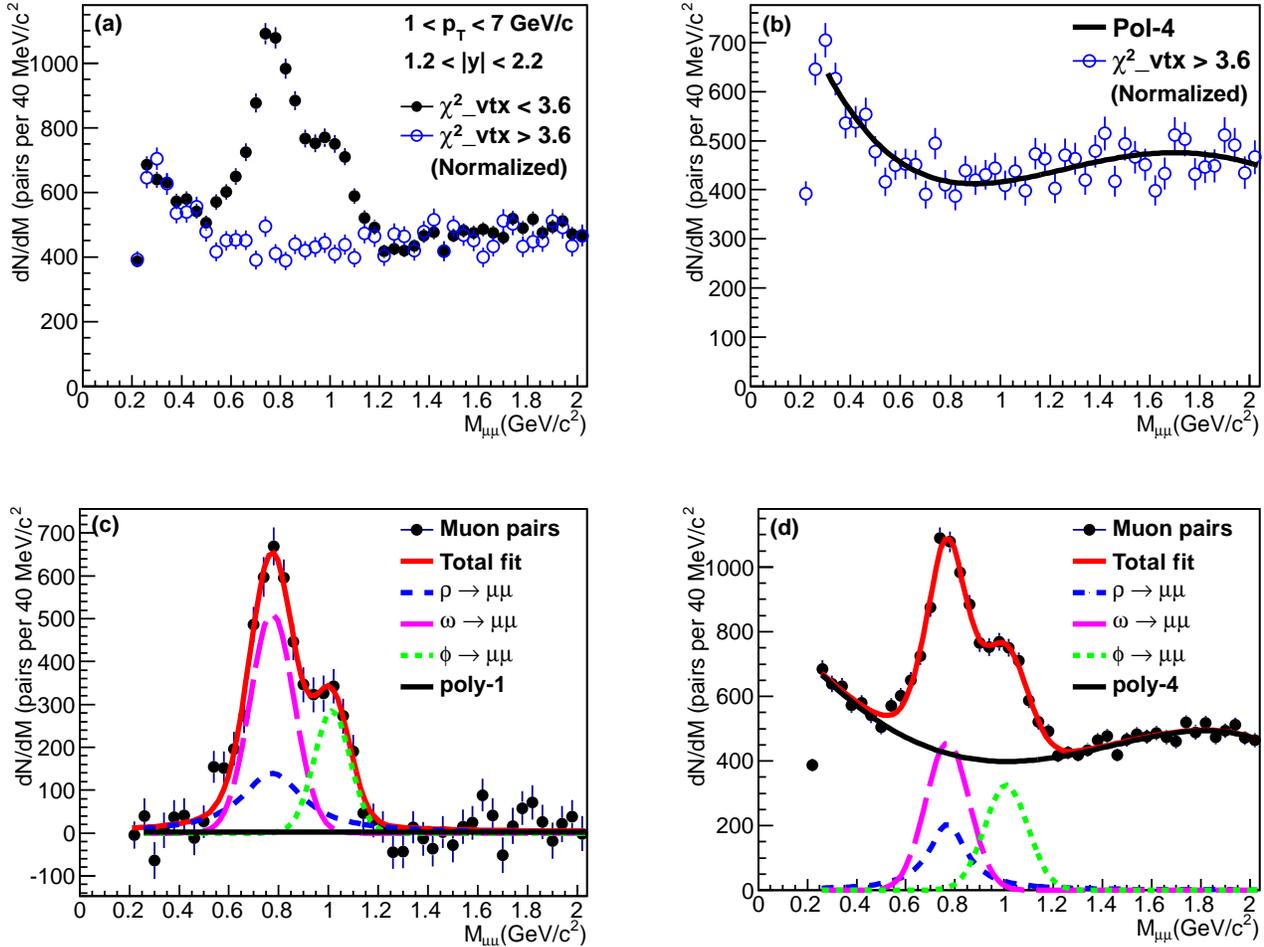}
\caption{\label{fig:northsigbkgdfit} (color online) 
Raw unlike-sign dimuon spectra (solid black circles) along with normalized 
background (empty blue circles) separated by $\chi^2_{\rm vtx,cut}$ in 
(a). Panel (b) shows the normalized background spectrum fitted with a 
fourth order polynomial. Panels (c) and (d) show the fitted spectra with 
(left) and without (right) background subtraction.
}
\end{figure*}

Figure~\ref{fig:northsigbkgdfit} shows an example of the different yield 
extraction methods.  Figure~\ref{fig:northsigbkgdfit}(a) shows the 
unlike-sign dimuon invariant mass spectrum (solid black circles) and the 
background spectrum (empty blue circles), while (b) shows the same 
background spectrum fitted with a fourth order polynomial. 
Figure~\ref{fig:northsigbkgdfit}(c) shows the unlike-sign dimuon invariant 
mass spectrum after subtracting the normalized background spectrum, shown 
in (b), fitted by two Gaussian distributions and a Breit Wigner 
distribution. s a cross check, a first order polynomial was added to the 
fit and the yields re-extracted and the resulting yields changed by less 
than 1\%. Figure~\ref{fig:northsigbkgdfit}(d) shows the unlike-sign 
dimuon invariant mass spectrum without background subtraction fitted by 
two Gaussian distributions, a Breit Wigner distribution and a fourth order 
polynomial constrained from the fit results shown in 
Fig.~\ref{fig:northsigbkgdfit}(b). The yields extracted using the two 
methods illustrated in Fig.~\ref{fig:northsigbkgdfit}(c) and (d) gave 
consistent results, well within uncertainties.

The data are binned as a function of \pt over the range $1<p_T<7$ 
GeV/$c$ for the rapidities $1.2<|y|<2.2$. In addition, the data 
integrated over the \pt range $1<p_T<7$ GeV/$c$ were studied as a 
function of rapidity. The raw yields in this measurement were extracted 
using background subtraction as well as background fit methods, and in the 
case of the background fit, several polynomials of different orders were 
attempted. As an example, the invariant mass spectra are fitted by the 
function that includes a fourth order polynomial, as defined below,

\begin{widetext}
\begin{equation}
f(x) = 0.58\times N_{\omega}\times BW(x,M_{\omega+\rho},\Gamma_{\rho}) + \frac{N_{\omega}}{\sqrt{2\pi}\sigma_{\omega}}G(x,M_{\omega+\rho},\sigma_{\omega}) + \frac{N_{\phi}}{\sqrt{2\pi}\sigma_{\phi}}G(x,M_{\phi},\sigma_{\phi}) + pol4
\end{equation}
\end{widetext}
where $BW$ and $G$ are a Breit-Wigner and a Gaussian functions, 
respectively, and $pol4$ is a fourth order polynomial. $N_\omega$ and 
$N_{\phi}$ are the yields of $\omega$ and $\phi$, and $M_{\omega+\rho}$ 
and $M_\phi$ are their mean values. The fit functions of $\omega$ 
(Gaussian) and $\rho$ (Breit Wigner) are constrained to have the same mean 
value and the ratio of their yields, $N_{\rho}/N_\omega$ is set to 0.58. 
The factor 0.58 is the ratio of $\rho$ and $\omega$ cross sections, 
$\sigma_\rho/\sigma_\omega = 1.15\pm0.15$~\cite{PhysRevC.81.034911}, 
multiplied by the ratio of their branching 
ratios~\cite{PhysRevD.86.010001}. The results of fitting the invariant 
mass spectra for different $p_{T}$ bins at $1.2<y<2.2$ are listed in 
Table~\ref{tab:fitPtRes}.

\begin{table*}[ht]
\caption{
The results of fitting the foreground spectrum by a function that includes 
two Gaussian distributions, a Breit Wigner distribution and a fourth order 
polynomial, over the mass range $0.3<M_{\mu\mu}<2.0$~GeV/$c^2$ for the 
listed \pt bins.
}
\begin{ruledtabular} \begin{tabular}{cccccc} 
 $p_T$ (GeV/$c$)               &  1.0 - 2.0      & 2.0 - 2.5             & 2.5 - 3.0             & 3.0 - 4.5             & 4.5 - 7.0 \\
\hline
 $N_\omega$        & (68 $\pm$ 5)$\times10^1$   & (63 $\pm$ 8)$\times10^1$     & (39 $\pm$ 4)$\times10^1$      & (36 $\pm$ 5)$\times10^1$      & (4.8 $\pm$ 1.2)$\times10^1$\\
 $M_{\omega+\rho}$ (GeV/$c^{2}$) & (77 $\pm$ 1)$\times10^{-2}$    & (77 $\pm$ 1)$\times10^{-2}$  & (77 $\pm$ 1)$\times10^{-2}$   & (76 $\pm$ 1)$\times10^{-2}$    & (80 $\pm$ 2)$\times10^{-2}$\\
 $\Gamma_{\rho}$ (GeV/$c^{2}$)   & (18 $\pm$ 4)$\times10^{-2}$    & (22 $\pm$ 4)$\times10^{-2}$  & (22 $\pm$ 2)$\times10^{-2}$   & (18 $\pm$ 4)$\times10^{-2}$    & (19 $\pm$ 2)$\times10^{-2}$\\
 $\sigma_{\omega}$ (GeV/$c^{2}$) & (8.8 $\pm$ 1.3)$\times10^{-2}$ & (85 $\pm$ 8)$\times10^{-3}$  & (8.8 $\pm$ 1.2)$\times10^{-2}$ & (8.1 $\pm$ 1.3)$\times10^{-2}$ & (7.2 $\pm$ 1.6)$\times10^{-2}$\\
 $N_{\phi}$        & (39  $\pm$ 8)$\times10^1$      & (53 $\pm$ 6)$\times10^1$    & (32 $\pm$ 4)$\times10^1$      & (28 $\pm$ 3)$\times10^1$       & 38 $\pm$ 10\\
 $M_{\phi}$ (GeV/$c^{2}$) & (100 $\pm$ 1)$\times10^{-2}$   & (99 $\pm$ 1)$\times10^{-2}$  & (100 $\pm$ 1)$\times10^{-2}$  & (100 $\pm$ 2)$\times10^{-2}$   & (106 $\pm$ 6)$\times10^{-2}$\\
 $\sigma_{\phi}$ (GeV/$c^{2}$)   & (7.5 $\pm$ 1.4)$\times10^{-2}$ & (8.8 $\pm$ 1.3)$\times10^{-2}$ & (8.8 $\pm$ 1.1)$\times10^{-2}$ & (8.8 $\pm$ 1.0)$\times10^{-2}$ & (7.2 $\pm$ 1.1)$\times10^{-2}$\\
 p0               & (20 $\pm$ 4)$\times10^1$       & (5.9 $\pm$ 3.8)$\times10^1$ & (13 $\pm$ 3)$\times10^1$   & (9.5 $\pm$ 2.8)$\times10^1$    & 8.6 $\pm$ 1.3\\
 p1               &(-3.8 $\pm$ 2.0)$\times10^2$    & (3.0 $\pm$ 1.8)$\times10^2$ &(-2.5 $\pm$ 1.3)$\times10^2$   &(-1.8 $\pm$ 1.3)$\times10^2$    &-15 $\pm$ 2.2\\
 p2               & (6.2 $\pm$ 3.1)$\times10^2$    &(-4.9 $\pm$ 2.5)$\times10^2$ & (3.4 $\pm$ 1.8)$\times10^2$   & (2.2 $\pm$ 1.8)$\times10^2$    & 39 $\pm$ 1.5\\
 p3               &(-3.6 $\pm$ 2.0)$\times10^2$    & (2.6 $\pm$ 1.4)$\times10^2$ &(-2.1 $\pm$ 1.0)$\times10^2$   &(-1.3 $\pm$ 1.0)$\times10^2$    &-35 $\pm$ 1\\
 p4               & (6.7 $\pm$ 4.3)$\times10^1$    &(-4.7 $\pm$ 2.9)$\times10^1$ & (4.6 $\pm$ 2.1)$\times10^1$   & (2.6 $\pm$ 2.1)$\times10^1$    & 9.2 $\pm$ 0.4\\
$\chi^2$/ndf     & 43.2/33                         & 28.1/33                     & 24.7/33                      & 29.2/33             & 39.7/33 \\
\end{tabular} \end{ruledtabular}
\label{tab:fitPtRes}
\end{table*}
 
The extracted yields of $\omega + \rho$ and $\phi$ were consistent among 
all fits. Therefore, the yields and their uncertainties of the fit with 
the best $\chi^2$ are used in the differential cross section calculations. 
The variations between the yields of the fit with the best $\chi^2$ and 
those of the other fits are considered as systematic uncertainties on the 
yield extraction.

The acceptance and reconstruction efficiency ($A\varepsilon_\mathrm{rec}$) 
of the muon spectrometers, including the MuID trigger efficiency, is 
determined by individually running {\sc pythia} 6.421 
(Default)~\cite{Phys.Commun.135.238} generated $\omega$, $\rho$, and 
$\phi$ through a full {\sc geant} simulation of the PHENIX detector. The 
simulated vertex distribution was tuned to match that of the 2009 data. 
The simulated events are reconstructed in the same manner as the data and 
the same cuts are applied as in the real data analysis.

\begin{figure}
\includegraphics[width=1.0\linewidth]{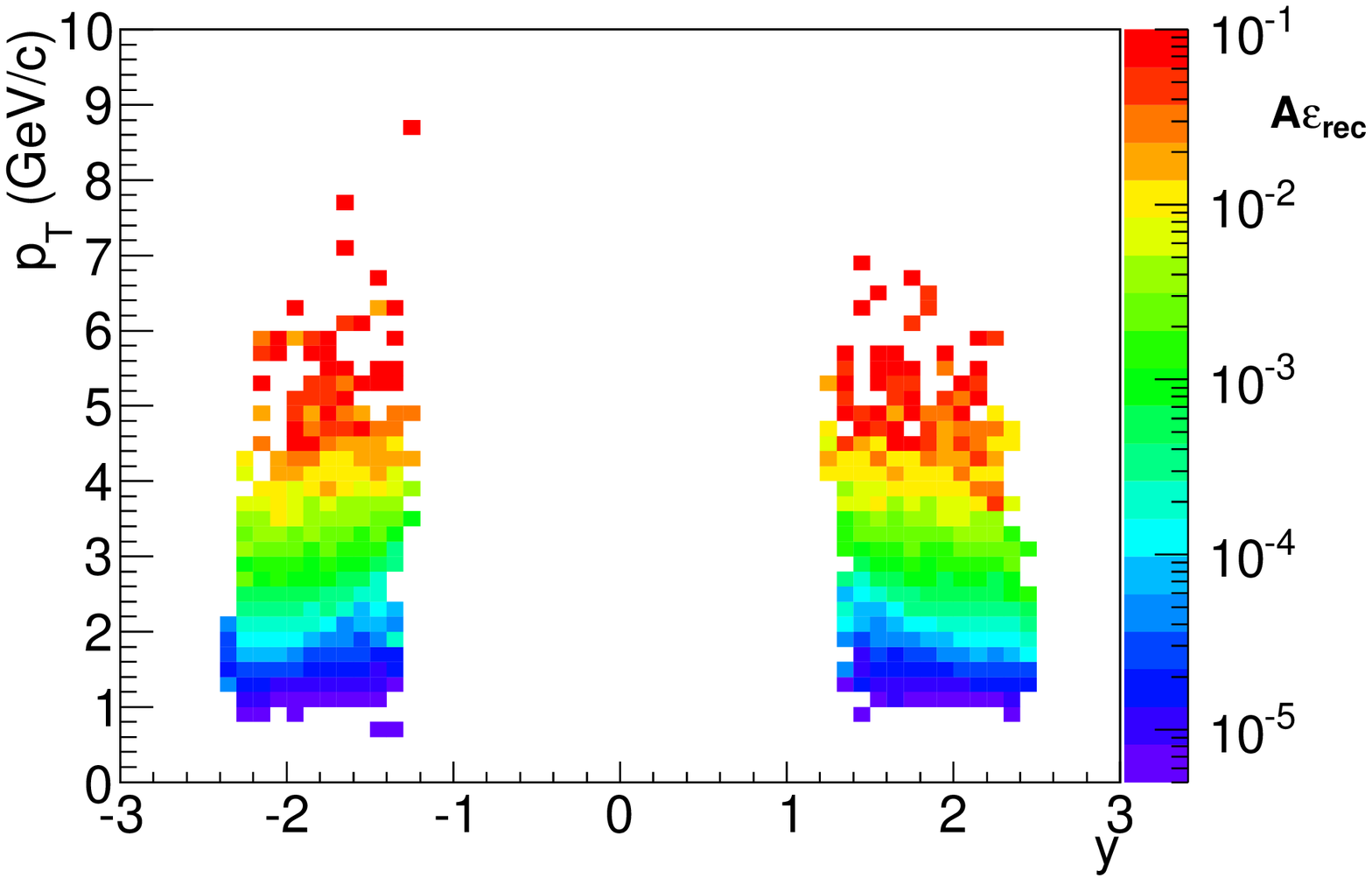}
\caption{\label{fig:AccEff} (color online) The $A\varepsilon_{\rm rec}$ as a function of rapidity ($x$-axis) and \pt ($y$-axis) for $\omega$.}
\end{figure}

The \pt and rapidity distributions of the generated events match the 
measured ones very well. The insert in Fig.~\ref{fig:bknorm} shows the 
$A\varepsilon_{\rm rec}$ as a function of invariant mass, while 
Fig.~\ref{fig:AccEff} shows the $A\varepsilon_{\rm rec}$ as a function of 
\pt and rapidity for $\omega$, as an example; the $A\varepsilon_{\rm rec}$ for 
$\rho$ and $\phi$ look very similar. The \pt dependent 
$A\varepsilon_{\rm rec}$ drops quickly at lower \pt which is the reason for 
limiting this study to $ p_T>1$ GeV/$c$.

\section{Results}

The differential cross section is evaluated according to the following 
relation:
\begin{equation}
BR \frac{d^2\sigma}{dydp_T} = \frac{1}{\Delta y\Delta p_T}\frac{N}{A\varepsilon_{\rm rec}\varepsilon_{\rm BBC}}\frac{\sigma_{\rm BBC}}{N^{\rm BBC}_{\rm MB}}
\end{equation}

where $\sigma_{\rm BBC}$ is the PHENIX BBC sampled cross section, $23.0 
\pm 2.2$ mb at $\sqrt{s} = 200$ GeV, which is determined from the van der 
Meer scan technique~\cite{PhysRevD.79.012003}. $BR$ is the branching ratio 
to dimuons ($BR(\omega\rightarrow\mu\mu) = (9.0{\pm}3.1)\times10^{-5}$, 
$BR(\rho\rightarrow\mu\mu) = (4.55{\pm}0.28)\times10^{-5}$, and 
$BR(\phi\rightarrow\mu\mu) = 
(2.87{\pm}0.19)\times10^{-4}$)~\cite{PhysRevD.86.010001}. 
$\varepsilon_\mathrm{BBC}$ = $0.795\pm 0.02$, is the BBC efficiency for hard 
scattering events~\cite{PhysRevLett.92.051802}. $N^{\rm BBC}_\mathrm{MB}$ is 
the number of MB events, and $N$ is the number of the observed mesons. In 
the \pt dependent study, the LVM yields were extracted for each arm 
separately and the weighted average of the two arms was used in the 
differential cross section calculations. $A\varepsilon_{\rm rec}$ is the 
acceptance and reconstruction efficiency.

The $\omega$ and $\rho$ yields are measured together and the \pt dependent 
and rapidity dependent differential cross sections are reported as 
$BR(\omega\rightarrow\mu\mu)\times d^2\sigma/ dydp_T(\omega) + 
BR(\rho\rightarrow\mu\mu)\times d^2\sigma/ dydp_T(\rho)$ and 
$BR(\omega\rightarrow\mu\mu)\times d\sigma/ dy(\omega) + 
BR(\rho\rightarrow\mu\mu)\times d\sigma/ dy(\rho)$, respectively, to 
minimize the contribution of uncertainties from branching ratios and total 
cross sections needed to calculate the absolute ($\omega + \rho$) 
differential cross section. The $A\varepsilon_{\rm rec}$ for $\omega + \rho$ 
is taken as the weighted average of the individual $A\varepsilon_{\rm rec}$, 
where the averaging is done based on $\omega$ and $\rho$ branching ratios.

The systematic uncertainties associated with this measurement can be 
divided into three categories based upon the effect each source has on the 
measured results. All uncertainties are reported as standard deviations. 
Type-A : point-to-point uncorrelated uncertainties allow the data points 
to move independently with respect to one another and are added in 
quadrature with statistical uncertainties, and include a 3\% signal 
extraction uncertainty. Type-B : point-to-point correlated uncertainties 
allow the data points to move coherently within the quoted range. These 
systematic uncertainties include a 4\% uncertainty from MuID tube 
efficiency and 2\% from MuTr overall efficiency. An 8\% uncertainty on 
the yield is assigned to account for a 2\% absolute momentum scale 
uncertainty, which was estimated by measuring the $J/\psi$ mass. A 9\% 
(7\%) uncertainty is assigned to the $-2.2<y<-1.2$ ($1.2<y<2.2$) 
rapidity due to the uncertainties in the $A\varepsilon_{\rm rec}$ 
determination method itself. The $A\varepsilon_{\rm rec}$ at the lowest \pt 
bin is small, as shown in~Fig.~\ref{fig:AccEff}, and sensitive to 
variations in the slope of the input \pt distribution which affects the 
differential cross section calculations at this \pt bin. To understand 
this effect, the \pt-dependent cross section is fitted by three commonly 
used fit functions (Hagedorn~\cite{NuovoCim.Suppl.3.147}, 
Kaplan~\cite{Phys.Rev.Lett.40.435}, and 
Tsallis~\cite{PHENIX_pp_200_central}) over the \pt range , $2<p_T<7$ 
GeV/$c$, and the fitted functions are extrapolated to lowest \pt bin, 
$1<p_T<2$ GeV/$c$. The differences between the values extracted from 
these fits and the measured one at the lowest \pt bin is within 8\%, 
hence an 8\% systematic uncertainty is assigned to lowest \pt bin to 
account for these differences. For the integrated and rapidity dependent 
cross sections the 8\% uncertainty is assigned to all data bins because 
the lowest \pt bin is dominant. Type-B systematic uncertainties are added 
in quadrature and are shown as shaded bands on the associated data points. 
Finally, an overall normalization uncertainty of 10\% was assigned for 
the BBC cross section and efficiency uncertainties which allows the data 
points to move together by a common multiplicative factor, and are labeled 
as type-C. These systematic uncertainties are listed in 
Table~\ref{tab:sysUncer}.

\begin{table}[ht]
\caption{
Systematic uncertainties included in the invariant yield and differential 
cross section calculations, where S (N) is for the $-2.2<y<-1.2$ 
($1.2<y<2.2$) rapidity. As explained in the text, there is an 8\% type-B 
systematic uncertainty due to small acceptance that impacts the low \pt 
region only which is not listed below.
}
\begin{ruledtabular} \begin{tabular}{ccc} 
 Type & Origin                        & Value (S/N)\\
\hline
  A   & Signal extraction             & 3\%\\
\\
  B   & MuID efficiency               & 4\%\\
  B   & MuTr efficiency               & 2\%\\
  B   & $A\varepsilon_{\rm rec}$           & 9\% / 7\%\\
  B   & Absolute momentum scale       & 8\%\\
Total & Quadratic sum of (B)        & 13\% / 12\%\\
\\
  C  & BBC  efficiency (Global)       & 10\%\\
\end{tabular} \end{ruledtabular}
\label{tab:sysUncer}
\end{table}

\begin{figure}[htb]
\includegraphics[width=1.0\linewidth]{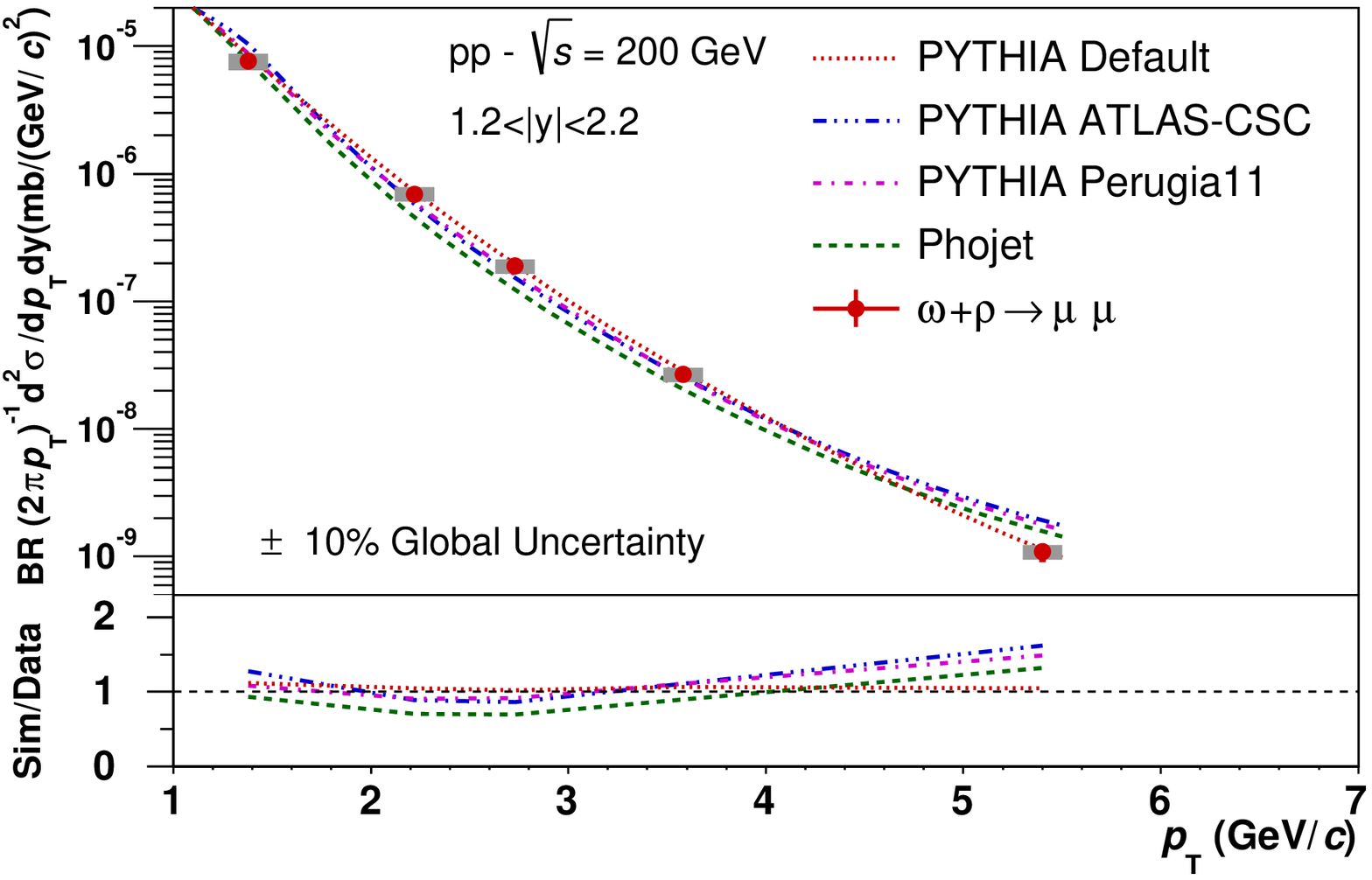}
\caption{\label{fig:ptXSomgRho} (color online) 
(top) \pt dependent differential cross sections of $\omega + \rho$ at 
rapidity, $1.2<|y|<2.2$. The error bars represent the statistical 
uncertainties, and the gray shaded band represents the quadratic sum of 
type-B systematic uncertainties. The data are compared with the {\sc 
pythia} {\sc atlas-csc}, default and {\sc prugia-11} tunes and 
{\sc phojet}. (bottom) Ratio between data and models. 
} 
\includegraphics[width=1.0\linewidth]{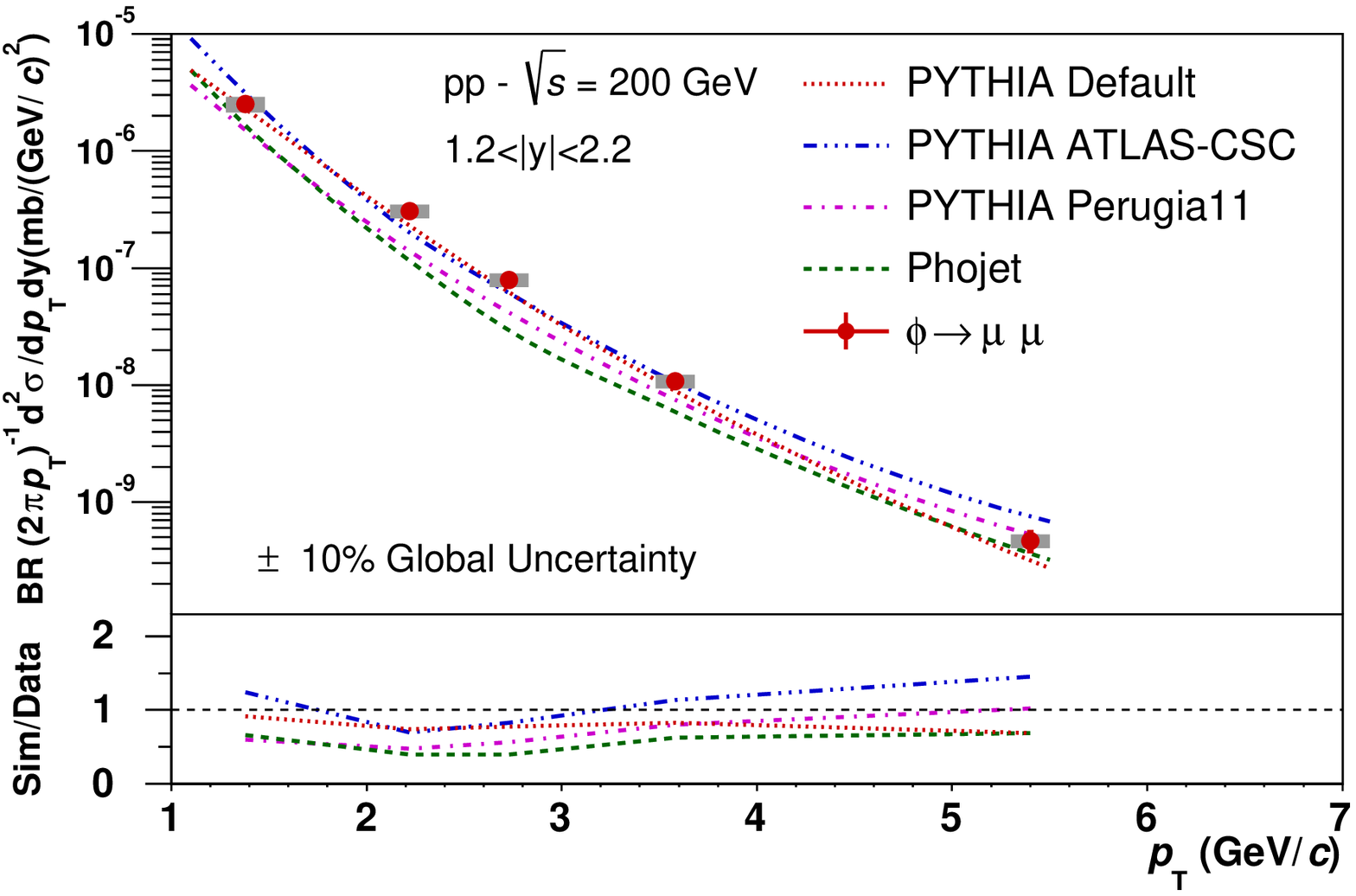}
\caption{\label{fig:ptXSphi} (color online) 
(top) \pt dependent differential cross sections of $\phi$ at rapidity, 
$1.2<|y|<2.2$. The error bars represent the statistical uncertainties, 
and the gray shaded band represents the quadratic sum of type-B systematic 
uncertainties. The data are compared with the {\sc pythia} 
({\sc atlas-csc}, default and {\sc perugia-11} tunes and {\sc phojet}. 
(bottom) Ratio between data and models.
}
\end{figure}

\begin{figure}[htb]
\includegraphics[width=1.0\linewidth]{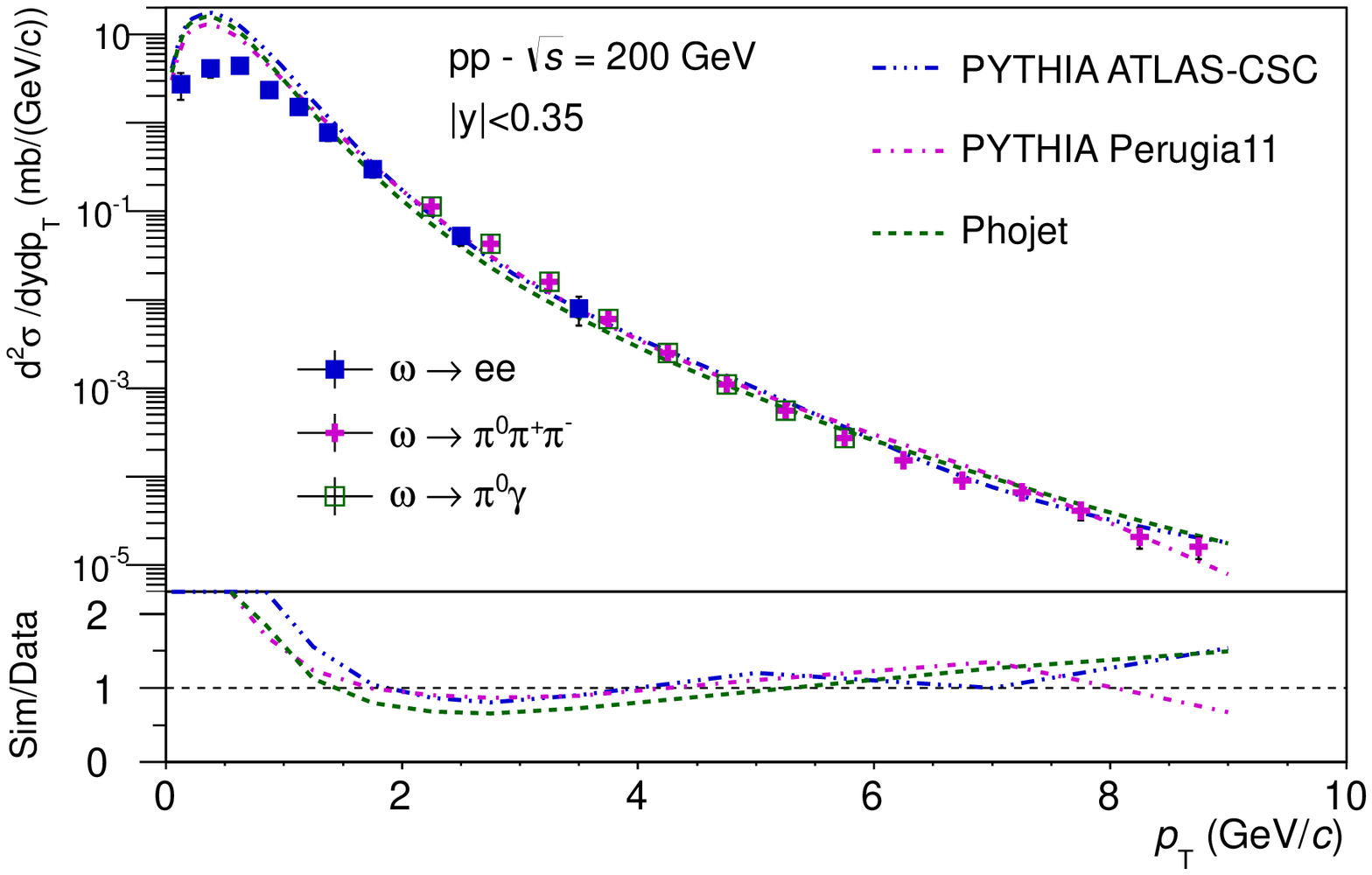}
\caption{\label{fig:ptXSomgRho_midrap} (color online) 
(top) \pt dependent differential cross sections of $\omega$ at rapidity, 
$|y|<0.35$~\cite{PHENIX_pp_200_central}. The data are compared with the 
{\sc pythia atlas-csc}, default and {\sc perugia-11} tunes and 
{\sc phojet}.  (bottom) Ratio between data and models.
}
\includegraphics[width=1.0\linewidth]{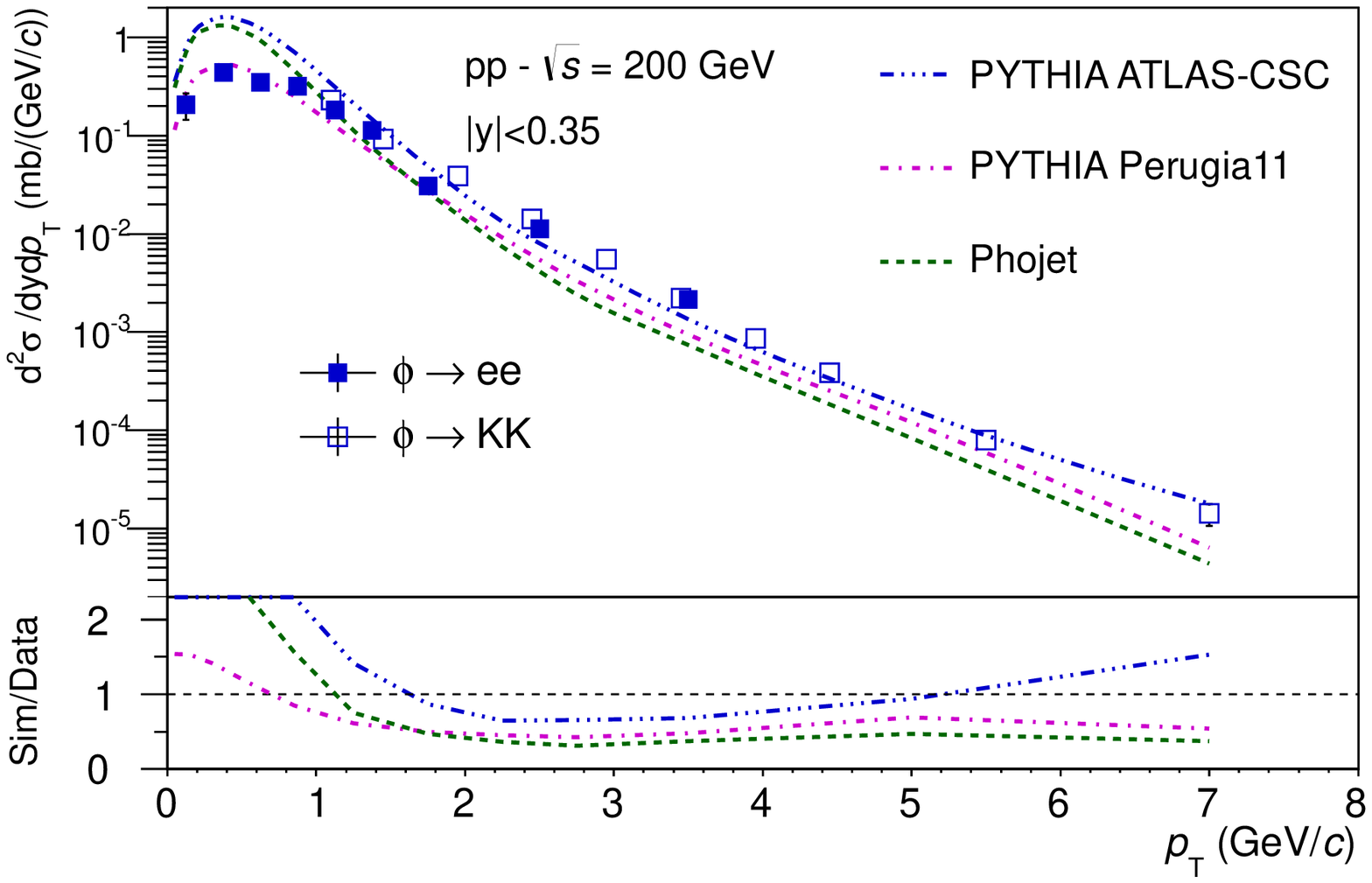}
\caption{\label{fig:ptXSphi_midrap} (color online) 
(top) \pt dependent differential cross sections of $\phi$ at rapidity, 
$|y|<0.35$~\cite{PHENIX_pp_200_central}. The data are compared with the 
{\sc pythia atlas-csc}, default and {\sc perugia-11} tunes and 
{\sc phojet}.  (bottom) Ratio between data and models.
}
\end{figure}

The open charm contribution to the signal is a possible source of 
systematic uncertainty. Even though the background subtracted dimuon 
spectrum in Fig.~\ref{fig:northsigbkgdfit}(c) shows no evidence of a 
remaining background, a Monte Carlo simulation was carried out to verify 
that the open charm contribution to the signal is negligible after 
background subtraction. A single particle {\sc pythia} simulation of open 
charm was generated and run through the PHENIX simulation chain. The charm 
differential cross section at forward rapidity, $d\sigma_{c\bar{c}}/dy|_{y 
= 1.6} = 0.243 \pm 0.013 (\mbox{stat}) \pm 0.105 (\mbox{data syst}) 
^{+0.049}_{-0.087} (\mbox{{\sc pythia} syst})$ 
mb~\cite{PhysRevD.76.092002}, is used with an inclusive branching ratio, 
$BR(D \rightarrow \mu + X)$ = 0.176~\cite{PhysRevD.86.010001}. The 
simulated events were then reconstructed using identical code to that used 
in the real data analysis, and after applying all cuts used in the 
analysis, the surviving rate of open charm was negligible in comparison to 
the low mass vector meson yields. Additionally, similar study of the 
$\eta$ and $\omega$ Dalitz decays showed that they were negligible in 
comparison to the low mass vector meson yields.

The differential cross sections for $\omega + \rho$ and $\phi$ as a 
function of \pt are shown in Figs.~\ref{fig:ptXSomgRho} 
and~\ref{fig:ptXSphi}, respectively, and listed in 
Table~\ref{tab:dxsdydpt}. The appropriate \pt value where each point was 
plotted is chosen such that the fit function, a function selected to fit 
the \pt distribution, is equal to its mean value~\cite{NIMA.355.541.1995} 
where the results are listed in the first column in 
Table~\ref{tab:dxsdydpt}.  Figs.~\ref{fig:ptXSomgRho} 
and~\ref{fig:ptXSphi} also 
include some standard tunes of 
{\sc pythia} ({\sc atlas-csc}~\cite{Acta.Phys.Pol.B.35.433}, 
default~\cite{Phys.Commun.135.238} and 
{\sc perugia-11}~\cite{PhysRevD.82.074018}) and 
{\sc phojet}~\cite{Phys.Rev.D.54.4244}. The bottom panels in 
Figs.~\ref{fig:ptXSomgRho}~and~\ref{fig:ptXSphi}
show the ratio between the measurement and the model predictions.

\begin{table*}[ht]
\caption{
Differential cross sections in b/(GeV/$c$) and \pt in (GeV/$c$) of $\omega 
+ \rho$ and $\phi$ at $1.2<|y|<2.2$ with statistical and type-A 
systematic uncertainties added in quadrature and type-B systematic 
uncertainties.
}
\begin{ruledtabular} \begin{tabular}{ccccc} 
$p_T$ & & $\frac{BR}{2\pi p_T}\frac{d^2\sigma_{\omega+\rho\rightarrow\mu\mu}}{dydp_T}$ & & $\frac{BR}{2\pi p_T}\frac{d^2\sigma_{\phi\rightarrow\mu\mu}}{dydp_T}$\\
(GeV/$c$) &  \smallskip &  (b / (GeV/$c$)$^2$) & \smallskip & (b / (GeV/$c$)$^2$) \\
\hline
1.38 & & (8.41 $\pm$ 0.67 $\pm$ 1.26)$\times10^{-09}$ & & (2.76 $\pm$ 0.35 $\pm$ 0.41)$\times10^{-09}$ \\
2.17 & & (7.19 $\pm$ 0.71 $\pm$ 0.93)$\times10^{-10}$ & & (3.19 $\pm$ 0.36 $\pm$ 0.41)$\times10^{-10}$ \\
2.65 & & (1.95 $\pm$ 0.19 $\pm$ 0.25)$\times10^{-10}$ & & (8.16 $\pm$ 0.93 $\pm$ 1.06)$\times10^{-11}$ \\
3.58 & & (2.68 $\pm$ 0.29 $\pm$ 0.35)$\times10^{-11}$ & & (1.09 $\pm$ 0.14 $\pm$ 0.14)$\times10^{-11}$ \\
5.40 & & (1.10 $\pm$ 0.16 $\pm$ 0.14)$\times10^{-12}$ & & (4.71 $\pm$ 0.90 $\pm$ 0.61)$\times10^{-13}$ \\
\end{tabular} \end{ruledtabular}
\label{tab:dxsdydpt}
\end{table*}

These model predictions were also tested against previously published 
midrapidity data~\cite{PHENIX_pp_200_central} as shown in 
Figs.~\ref{fig:ptXSomgRho_midrap} and~\ref{fig:ptXSphi_midrap}.

{\sc pythia atlas-csc} and {\sc perugia-11} tunes, reproduce the 
differential cross section at both midrapidity and forward rapidity for 
$\omega$ and $\omega+\rho$, respectively, while {\sc phojet} under 
predicts the data in both cases. The {\sc pythia atlas-csc} reproduces the 
$\phi$ differential cross sections at forward rapidities. The {\sc pythia 
atlas-csc} and {\sc perugia-11} tunes and {\sc phojet} fail to match the 
data below 1 GeV/$c$. Generally, {\sc pythia} and {\sc phojet} seem to do 
better job reproducing $\omega+\rho$ than $\phi$.

\begin{figure}[htb]
\includegraphics[width=1.0\linewidth]{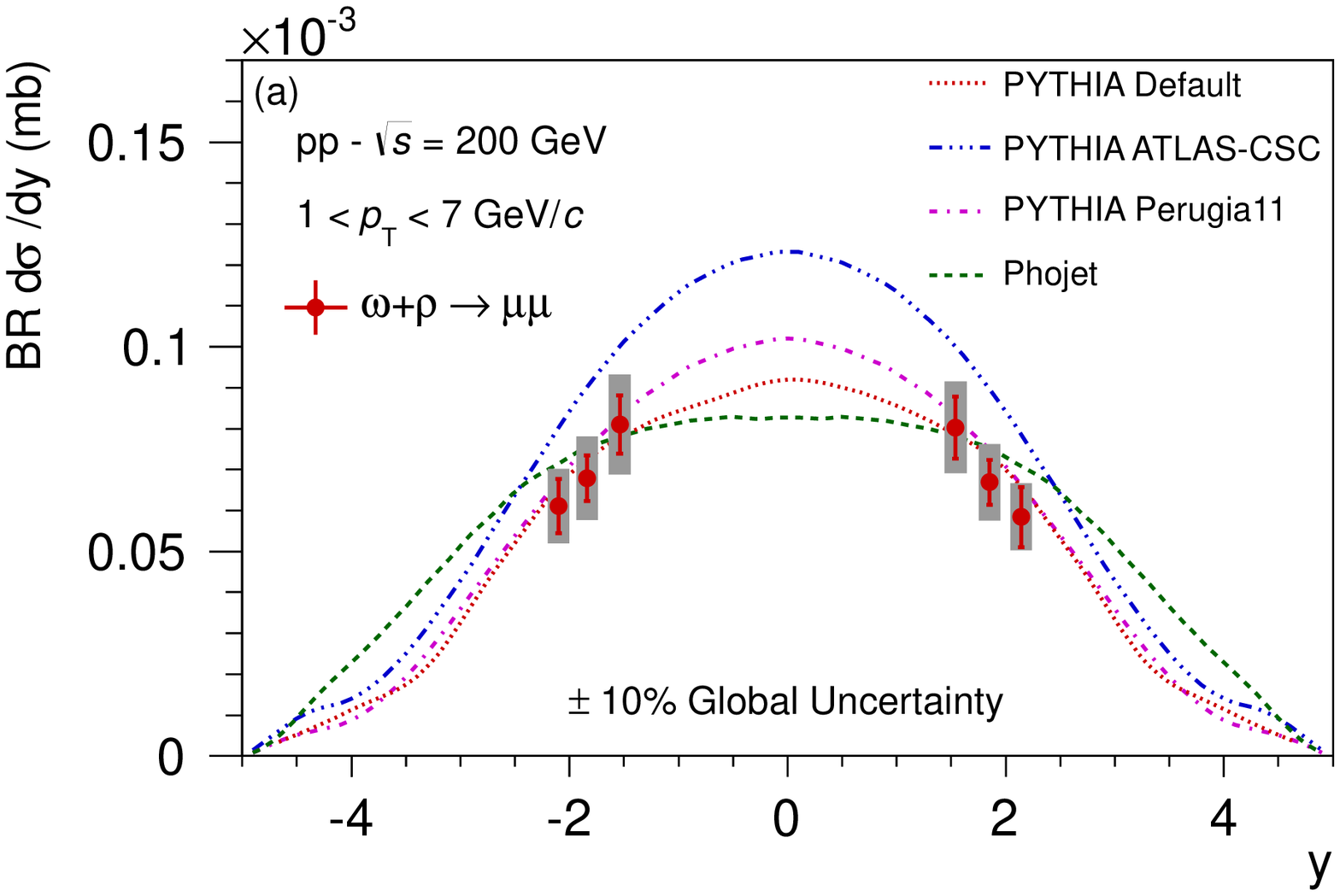}
\includegraphics[width=1.0\linewidth]{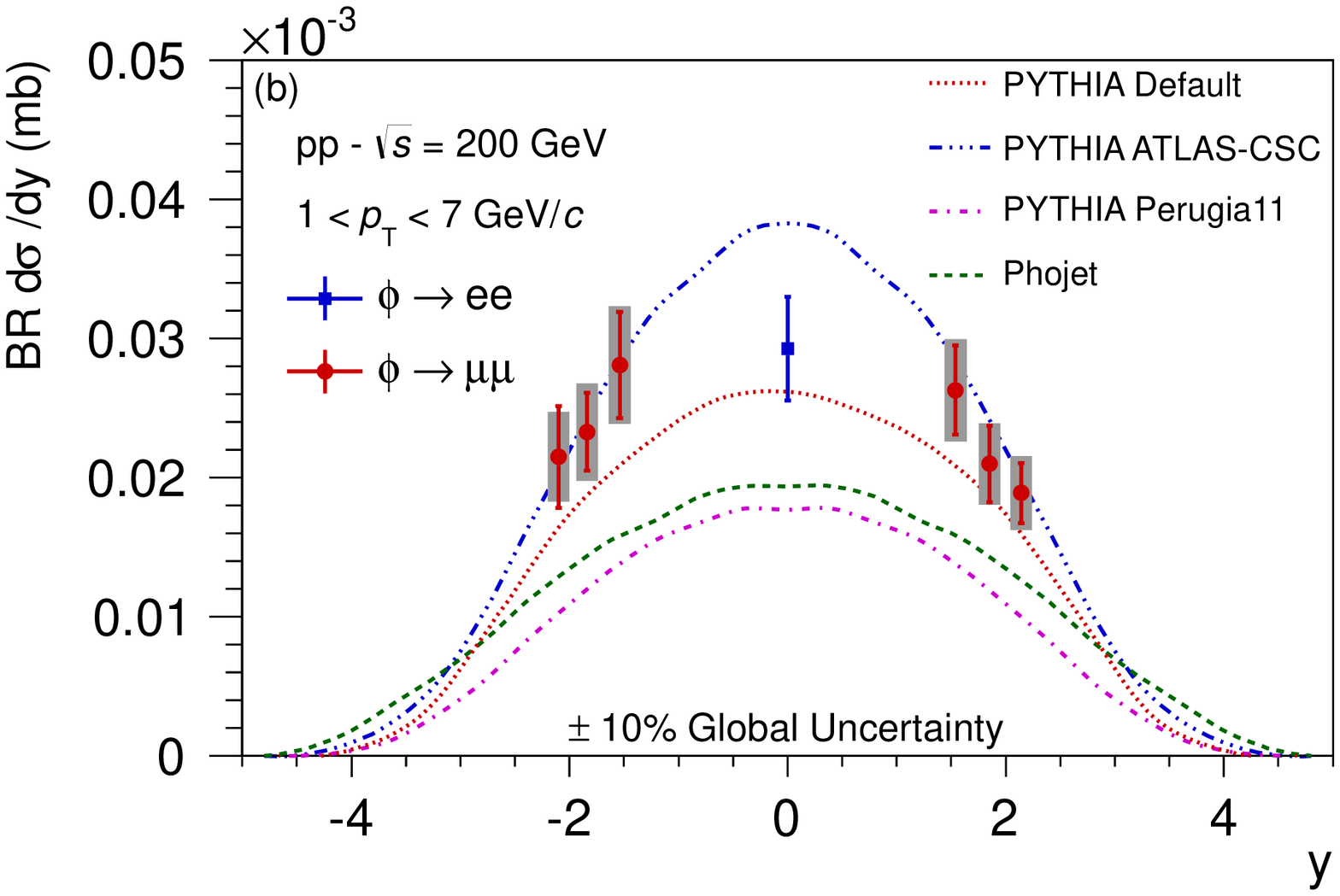}
\caption{\label{fig:rapXS} (color online) 
Rapidity dependent differential cross section of $\omega + \rho$ (a) and 
$\phi$ (b) along with previous PHENIX results~\cite{PHENIX_pp_200_central} 
summed over the \pt range, $1<\pt<7$ GeV. The error bars represent the 
quadratic sum of the statistical uncertainties and type-A systematic 
uncertainties, and the gray shaded band represents the quadratic sum of 
type-B systematic uncertainties. The data are compared with the 
{\sc pythia atlas-csc} and {\sc perugia-11} tunes and {\sc phojet}.
}
\end{figure}

\begin{table}[ht]
\caption{
Differential cross sections in b and rapidity of $\omega + \rho$ and 
$\phi$ at $1<p_T<7$ GeV/$c$ with statistical and type-A systematic 
uncertainties added in quadrature and type-B systematic uncertainties.
}
\begin{ruledtabular} \begin{tabular}{cccc} 
$y$ & $BR\frac{d\sigma_{\omega+\rho\rightarrow\mu\mu}}{dy}$ (nb) & &  $BR\frac{d\sigma_{\phi\rightarrow\mu\mu}}{dy}$ (nb)\\
\hline
-2.10 & 61.1 $\pm$ 6.7 $\pm$ 9.2  & &  21.5 $\pm$ 3.7 $\pm$ 3.2\\
-1.84 & 67.9 $\pm$ 5.6 $\pm$ 10.2 & &  23.3 $\pm$ 2.8 $\pm$ 3.5\\
-1.54 & 81.0 $\pm$ 7.1 $\pm$ 12.2 & &  28.1 $\pm$ 3.8 $\pm$ 4.2\\
 1.54 & 80.3 $\pm$ 7.6 $\pm$ 11.2 & &  26.3 $\pm$ 3.2 $\pm$ 3.7\\
 1.85 & 66.9 $\pm$ 5.4 $\pm$ 9.4  & &  21.0 $\pm$ 2.8 $\pm$ 2.9\\
 2.14 & 58.4 $\pm$ 7.4 $\pm$ 8.2  & &  18.9 $\pm$ 2.2 $\pm$ 2.6\\
\end{tabular} \end{ruledtabular}
\label{tab:dxsdy}
\end{table}

\begin{table}[ht]
\caption{
$N_\phi/(N_\omega+N_\rho)$ and \pt in (GeV/$c$) with statistical and 
type-A systematic uncertainties added in quadrature and type-B systematic 
uncertainties.
}
\begin{ruledtabular} \begin{tabular}{cc} 
$p_T$ (GeV/$c$) & $N_\phi/(N_\omega+N_\rho)$  \\
\hline
1.38 & 0.33 $\pm$ 0.04 $\pm$ 0.03\\
2.17 & 0.44 $\pm$ 0.05 $\pm$ 0.04\\
2.65 & 0.43 $\pm$ 0.05 $\pm$ 0.04\\
3.58 & 0.40 $\pm$ 0.05 $\pm$ 0.04\\
5.40 & 0.45 $\pm$ 0.09 $\pm$ 0.04\\
\end{tabular} \end{ruledtabular}
\label{tab:yieldRat}
\end{table}

Figure~\ref{fig:rapXS} and Table~\ref{tab:dxsdy} show the differential 
cross section as a function of rapidity for $\omega + \rho$ in (a) and 
$\phi$ in (b), along with {\sc pythia} tunes 
({\sc atlas-csc}, default, and {perugia-11}) 
and {\sc phojet}. It can be seen in Fig.~\ref{fig:rapXS} that default 
and {sc perugia-11} tunes reproduce the $\omega + \rho$ results, while the 
{\sc atlas-csc} tune matches the $\phi$ forward rapidity results.

The acceptance at low \pt is very small to negligible in the low mass 
region which prevents us from extracting the differential cross sections, 
$d\sigma/dy$, summed over all \pt directly from the data. Instead, we 
report $d\sigma/dy$ integrated over the measured \pt range, 
$d\sigma/dy(\omega+\rho\rightarrow \mu\mu) (1<p_T<7\mbox{ GeV/$c$}, 
1.2<|y|<2.2) = 80 \pm 6 \mbox{ (stat)} \pm 12 \mbox{ (syst)}$ nb 
and $d\sigma/dy(\phi\rightarrow \mu\mu) (1<p_T<7 \mbox{ GeV/$c$}, 
1.2<|y|<2.2)=27 \pm 3 \mbox{ (stat)} \pm 4 \mbox{ (syst)} $ nb.

\begin{figure}[htb]
\includegraphics[width=1.0\linewidth]{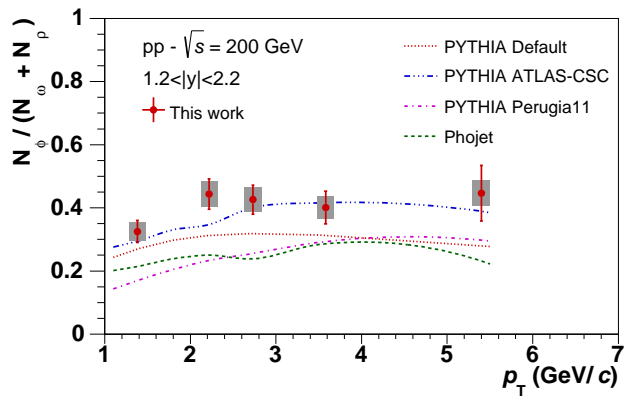}
\caption{\label{fig:yieldRat} (color online)  
$N_\phi/(N_\omega+N_\rho)$ as a function of \pt. The error bars represent 
the quadratic sum of the statistical uncertainties and type-A systematic 
uncertainties, and the gray shaded band represents the quadratic sum of 
type-B systematic uncertainties.
}
\end{figure}

The ratio $N_\phi/(N_\omega+N_\rho) = 
BR(\phi\rightarrow\mu\mu)\sigma_\phi/(BR(\omega\rightarrow\mu\mu)\sigma_\omega 
+ BR(\rho\rightarrow\mu\mu)\sigma_\rho)$, corrected for acceptance and 
efficiency, was determined for $1<p_T<7$ GeV/$c$ and $1.2<|y|<2.2$, 
giving $0.390 \pm 0.021 \mbox{ (stat)} \pm 0.035 \mbox{ (syst)}$, 
as shown in Fig.~\ref{fig:yieldRat} and listed in 
Table~\ref{tab:yieldRat}. Systematic uncertainties including MuID and MuTr 
efficiencies, absolute momentum scale and BBC efficiency cancel out when 
taking the yield ratio.

Figure~\ref{fig:yieldRat} also shows {\sc pythia} ({\sc atlas-csc}, 
default, and {sc perugia-11} tunes) and {\sc phojet}. The {\sc atlas-csc} 
tune reproduces the ratio while the other models underestimate it. The 
ALICE experiment also measured this ratio in \pp collisions at $\sqrt{s}=$ 
7 TeV in the dimuon rapidity region $2.5<y<4$. The reported value is 
$0.416 \pm 0.032 \mbox{ (stat)} \pm 0.004 \mbox{ (syst)}$~\cite{ALICE_pp_forward} 
over the \pt range $1<p_T<5$ which is consistent with our result.

\section{Summary and Conclusions}

In summary, we studied the low mass vector meson, $\omega$, $\rho$, and 
$\phi$, production in \pp collisions at $\sqrt{s}$= 200 GeV for 
$1.2<|y|<2.2$ and $1.0<p_T<7.0$ GeV/$c$, through the dimuon decay channel. 
We measured $\omega+\rho$, and $\phi$ differential cross sections as a 
function of \pt as well as a function of rapidity.

The differential cross sections, $d\sigma/dy$ of $\omega+\rho$ and $\phi$, 
were evaluated over the measured \pt range, 
$d\sigma/dy(\omega+\rho\rightarrow \mu\mu) (1<p_T<7 \mbox{ GeV/$c$}, 
1.2<|y|<2.2) = 80 \pm 6 \mbox{ (stat)} \pm 12 \mbox{ (syst)}$~nb 
and $d\sigma/dy(\phi\rightarrow \mu\mu) (1<p_T<7 \mbox{ GeV/$c$}, 
1.2<|y|<2.2) = 27 \pm 3 \mbox{ (stat)} \pm 4 \mbox{ (syst)}$~nb. The 
ratio $N_\phi/(N_\omega+N_\rho)$, at $1<p_T<7$ GeV/$c$ and 
$1.2<|y|<2.2$, was also determined, and is $0.390 \pm 0.021 \mbox{ (stat)} \pm 
0.035 \mbox{ (syst)}$, which is consistent with ALICE measurement at 
larger rapidity and higher energy. This agreement with the ALICE result at 
$\sim0.4$, which is higher than {\sc pythia} default at $\sim0.3$, suggests a 
higher $g+g$ contribution to $\phi$ production.

The data are compared to some commonly used {\sc pythia} tunes and {\sc 
phojet}. Overall, the {\sc pythia atlas-csc} and default tunes describe 
forward rapidity data except for the $\phi$ rapidity distribution and 
describe midrapidity data above 1 GeV/$c$. The {\sc pythia perugia-11} 
tune describes the $\omega+\rho$ differential cross section while it 
underestimates the $\phi$ differential cross section. Generally, all these 
event generators describe the shape of the LVM \pt distribution indicating 
that leading-order perturbative QCD-based event generators can describe 
\pt distribution.


\section*{ACKNOWLEDGMENTS}


We thank the staff of the Collider-Accelerator and Physics
Departments at Brookhaven National Laboratory and the staff of
the other PHENIX participating institutions for their vital
contributions.  We acknowledge support from the 
Office of Nuclear Physics in the
Office of Science of the Department of Energy,
the National Science Foundation, 
Abilene Christian University Research Council, 
Research Foundation of SUNY, and
Dean of the College of Arts and Sciences, Vanderbilt University 
(U.S.A),
Ministry of Education, Culture, Sports, Science, and Technology
and the Japan Society for the Promotion of Science (Japan),
Conselho Nacional de Desenvolvimento Cient\'{\i}fico e
Tecnol{\'o}gico and Funda\c c{\~a}o de Amparo {\`a} Pesquisa do
Estado de S{\~a}o Paulo (Brazil),
Natural Science Foundation of China (P.~R.~China),
Ministry of Science, Education, and Sports (Croatia),
Ministry of Education, Youth and Sports (Czech Republic),
Centre National de la Recherche Scientifique, Commissariat
{\`a} l'{\'E}nergie Atomique, and Institut National de Physique
Nucl{\'e}aire et de Physique des Particules (France),
Bundesministerium f\"ur Bildung und Forschung, Deutscher
Akademischer Austausch Dienst, and Alexander von Humboldt Stiftung (Germany),
OTKA NK 101 428 grant and the Ch. Simonyi Fund (Hungary),
Department of Atomic Energy and Department of Science and Technology (India), 
Israel Science Foundation (Israel), 
National Research Foundation of Korea of the Ministry of Science,
ICT, and Future Planning (Korea),
Physics Department, Lahore University of Management Sciences (Pakistan),
Ministry of Education and Science, Russian Academy of Sciences,
Federal Agency of Atomic Energy (Russia),
VR and Wallenberg Foundation (Sweden), 
the U.S. Civilian Research and Development Foundation for the
Independent States of the Former Soviet Union, 
the Hungarian American Enterprise Scholarship Fund,
and the US-Israel Binational Science Foundation.



\begin{thebibliography}{32}
\expandafter\ifx\csname natexlab\endcsname\relax\def\natexlab#1{#1}\fi
\expandafter\ifx\csname bibnamefont\endcsname\relax
  \def\bibnamefont#1{#1}\fi
\expandafter\ifx\csname bibfnamefont\endcsname\relax
  \def\bibfnamefont#1{#1}\fi
\expandafter\ifx\csname citenamefont\endcsname\relax
  \def\citenamefont#1{#1}\fi
\expandafter\ifx\csname url\endcsname\relax
  \def\url#1{\texttt{#1}}\fi
\expandafter\ifx\csname urlprefix\endcsname\relax\def\urlprefix{URL }\fi
\providecommand{\bibinfo}[2]{#2}
\providecommand{\eprint}[2][]{\url{#2}}

\bibitem[{\citenamefont{Abelev et~al.}(2009{\natexlab{a}})}]{STAR_Phi2}
\bibinfo{author}{\bibfnamefont{B.}~\bibnamefont{Abelev}} \bibnamefont{et~al.}
  (\bibinfo{collaboration}{STAR Collaboration}), \bibinfo{journal}{Phys. Rev.
  C} \textbf{\bibinfo{volume}{79}}, \bibinfo{pages}{064903}
  (\bibinfo{year}{2009}{\natexlab{a}}).

\bibitem[{\citenamefont{Adare
  et~al.}(2011{\natexlab{a}})}]{PHENIX_pp_200_central}
\bibinfo{author}{\bibfnamefont{A.}~\bibnamefont{Adare}} \bibnamefont{et~al.}
  (\bibinfo{collaboration}{PHENIX Collaboration}), \bibinfo{journal}{Phys. Rev.
  D} \textbf{\bibinfo{volume}{83}}, \bibinfo{pages}{052004}
  (\bibinfo{year}{2011}{\natexlab{a}}).

\bibitem[{\citenamefont{Alexopoulos et~al.}(1995)}]{E735_ppbar_phi}
\bibinfo{author}{\bibfnamefont{T.}~\bibnamefont{Alexopoulos}}
  \bibnamefont{et~al.} (\bibinfo{collaboration}{E735 Collaboration}),
  \bibinfo{journal}{Z. Phys. C} \textbf{\bibinfo{volume}{67}},
  \bibinfo{pages}{411} (\bibinfo{year}{1995}).

\bibitem[{\citenamefont{Abelev
  et~al.}(2012{\natexlab{a}})}]{ALICE_pp_Phi_central}
\bibinfo{author}{\bibfnamefont{B.}~\bibnamefont{Abelev}} \bibnamefont{et~al.}
  (\bibinfo{collaboration}{ALICE Collaboration}), \bibinfo{journal}{Euro. Phys.
  Jour. C} \textbf{\bibinfo{volume}{72}}, \bibinfo{pages}{2183}
  (\bibinfo{year}{2012}{\natexlab{a}}).

\bibitem[{\citenamefont{Abelev et~al.}(2012{\natexlab{b}})}]{ALICE_pp_forward}
\bibinfo{author}{\bibfnamefont{B.}~\bibnamefont{Abelev}} \bibnamefont{et~al.}
  (\bibinfo{collaboration}{ALICE Collaboration}), \bibinfo{journal}{Phys. Lett.
  B} \textbf{\bibinfo{volume}{710}}, \bibinfo{pages}{557}
  (\bibinfo{year}{2012}{\natexlab{b}}).

\bibitem[{\citenamefont{Aaij et~al.}(2011)}]{LHCb_pp_Phi}
\bibinfo{author}{\bibfnamefont{R.}~\bibnamefont{Aaij}} \bibnamefont{et~al.}
  (\bibinfo{collaboration}{LHCb Collaboration}), \bibinfo{journal}{Phys. Lett.
  B} \textbf{\bibinfo{volume}{703}}, \bibinfo{pages}{267}
  (\bibinfo{year}{2011}).

\bibitem[{\citenamefont{Koch et~al.}(1986)\citenamefont{Koch, M\"{u}ller, and
  Rafelski}}]{Koch}
\bibinfo{author}{\bibfnamefont{P.}~\bibnamefont{Koch}},
  \bibinfo{author}{\bibfnamefont{B.}~\bibnamefont{M\"{u}ller}},
  \bibnamefont{and} \bibinfo{author}{\bibfnamefont{J.}~\bibnamefont{Rafelski}},
  \bibinfo{journal}{Phys. Rpts.} \textbf{\bibinfo{volume}{142}},
  \bibinfo{pages}{167} (\bibinfo{year}{1986}).

\bibitem[{\citenamefont{Alt et~al.}(2008)}]{NA49_phi}
\bibinfo{author}{\bibfnamefont{C.}~\bibnamefont{Alt}} \bibnamefont{et~al.}
  (\bibinfo{collaboration}{NA49 Collaboration}), \bibinfo{journal}{Phys. Rev.
  C} \textbf{\bibinfo{volume}{78}}, \bibinfo{pages}{044907}
  (\bibinfo{year}{2008}).

\bibitem[{\citenamefont{Alessandro et~al.}(2003)}]{NA50_Phi}
\bibinfo{author}{\bibfnamefont{B.}~\bibnamefont{Alessandro}}
  \bibnamefont{et~al.} (\bibinfo{collaboration}{NA50 Collaboration}),
  \bibinfo{journal}{Phys. Lett. B} \textbf{\bibinfo{volume}{555}},
  \bibinfo{pages}{147} (\bibinfo{year}{2003}).

\bibitem[{\citenamefont{Adamov\'a et~al.}(2006)}]{CERES_Phi}
\bibinfo{author}{\bibfnamefont{D.}~\bibnamefont{Adamov\'a}}
  \bibnamefont{et~al.} (\bibinfo{collaboration}{CERES Collaboration}),
  \bibinfo{journal}{Phys. Rev. Lett.} \textbf{\bibinfo{volume}{96}},
  \bibinfo{pages}{152301} (\bibinfo{year}{2006}).

\bibitem[{\citenamefont{Arnaldi et~al.}(2011)}]{NA60_Phi}
\bibinfo{author}{\bibfnamefont{R.}~\bibnamefont{Arnaldi}} \bibnamefont{et~al.}
  (\bibinfo{collaboration}{NA60 Collaboration}), \bibinfo{journal}{Phys. Lett.
  B} \textbf{\bibinfo{volume}{699}}, \bibinfo{pages}{325}
  (\bibinfo{year}{2011}).

\bibitem[{\citenamefont{Adare
  et~al.}(2011{\natexlab{b}})}]{PHENIX_Phi_RAA_200_central}
\bibinfo{author}{\bibfnamefont{A.}~\bibnamefont{Adare}} \bibnamefont{et~al.}
  (\bibinfo{collaboration}{PHENIX Collaboration}), \bibinfo{journal}{Phys. Rev.
  C} \textbf{\bibinfo{volume}{83}}, \bibinfo{pages}{024909}
  (\bibinfo{year}{2011}{\natexlab{b}}).

\bibitem[{\citenamefont{Abelev et~al.}(2009{\natexlab{b}})}]{STAR_Phi}
\bibinfo{author}{\bibfnamefont{B.}~\bibnamefont{Abelev}} \bibnamefont{et~al.}
  (\bibinfo{collaboration}{STAR Collaboration}), \bibinfo{journal}{Phys. Lett.
  B} \textbf{\bibinfo{volume}{673}}, \bibinfo{pages}{183}
  (\bibinfo{year}{2009}{\natexlab{b}}).

\bibitem[{\citenamefont{van Hees and Rapp}(2008)}]{vanHees}
\bibinfo{author}{\bibfnamefont{H.}~\bibnamefont{van Hees}} \bibnamefont{and}
  \bibinfo{author}{\bibfnamefont{R.}~\bibnamefont{Rapp}},
  \bibinfo{journal}{Nucl. Phys. A} \textbf{\bibinfo{volume}{806}},
  \bibinfo{pages}{339} (\bibinfo{year}{2008}).

\bibitem[{\citenamefont{Adamov\'a et~al.}(2008)}]{CERES}
\bibinfo{author}{\bibfnamefont{D.}~\bibnamefont{Adamov\'a}}
  \bibnamefont{et~al.} (\bibinfo{collaboration}{CERES Collaboration}),
  \bibinfo{journal}{Phys. Lett. B} \textbf{\bibinfo{volume}{666}},
  \bibinfo{pages}{425} (\bibinfo{year}{2008}).

\bibitem[{\citenamefont{Arnaldi et~al.}(2006)}]{NA60_Rho}
\bibinfo{author}{\bibfnamefont{R.}~\bibnamefont{Arnaldi}} \bibnamefont{et~al.}
  (\bibinfo{collaboration}{NA60 Collaboration}), \bibinfo{journal}{Phys. Rev.
  Lett.} \textbf{\bibinfo{volume}{96}}, \bibinfo{pages}{162302}
  (\bibinfo{year}{2006}).

\bibitem[{\citenamefont{Adcox et~al.}(2003)}]{NIMA.499.469}
\bibinfo{author}{\bibfnamefont{K.}~\bibnamefont{Adcox}} \bibnamefont{et~al.}
  (\bibinfo{collaboration}{PHENIX Collaboration}), \bibinfo{journal}{Nucl.
  Instrum. Methods A} \textbf{\bibinfo{volume}{499}}, \bibinfo{pages}{469}
  (\bibinfo{year}{2003}).

\bibitem[{\citenamefont{Aronson et~al.}(2003)}]{NIMA.499.480}
\bibinfo{author}{\bibfnamefont{S.}~\bibnamefont{Aronson}} \bibnamefont{et~al.}
  (\bibinfo{collaboration}{PHENIX Collaboration}), \bibinfo{journal}{Nucl.
  Instrum. Methods A} \textbf{\bibinfo{volume}{499}}, \bibinfo{pages}{480}
  (\bibinfo{year}{2003}).

\bibitem[{\citenamefont{Adare et~al.}(2011{\natexlab{c}})}]{PhysRevC.84.054912}
\bibinfo{author}{\bibfnamefont{A.}~\bibnamefont{Adare}} \bibnamefont{et~al.}
  (\bibinfo{collaboration}{PHENIX Collaboration}), \bibinfo{journal}{Phys. Rev.
  C} \textbf{\bibinfo{volume}{84}}, \bibinfo{pages}{054912}
  (\bibinfo{year}{2011}{\natexlab{c}}).

\bibitem[{\citenamefont{Adare et~al.}(2013)}]{PhysRevC.87.034904}
\bibinfo{author}{\bibfnamefont{A.}~\bibnamefont{Adare}} \bibnamefont{et~al.}
  (\bibinfo{collaboration}{PHENIX Collaboration}), \bibinfo{journal}{Phys. Rev.
  C} \textbf{\bibinfo{volume}{87}}, \bibinfo{pages}{034904}
  (\bibinfo{year}{2013}).

\bibitem[{\citenamefont{Adare et~al.}(2010)}]{PhysRevC.81.034911}
\bibinfo{author}{\bibfnamefont{A.}~\bibnamefont{Adare}} \bibnamefont{et~al.}
  (\bibinfo{collaboration}{PHENIX Collaboration}), \bibinfo{journal}{Phys. Rev.
  C} \textbf{\bibinfo{volume}{81}}, \bibinfo{pages}{034911}
  (\bibinfo{year}{2010}).

\bibitem[{\citenamefont{Beringer et~al.}(2012)}]{PhysRevD.86.010001}
\bibinfo{author}{\bibfnamefont{J.}~\bibnamefont{Beringer}} \bibnamefont{et~al.}
  (\bibinfo{collaboration}{Particle Data Group}), \bibinfo{journal}{Phys. Rev.
  D} \textbf{\bibinfo{volume}{86}}, \bibinfo{pages}{010001}
  (\bibinfo{year}{2012}).

\bibitem[{\citenamefont{Sjostrand et~al.}(2001)}]{Phys.Commun.135.238}
\bibinfo{author}{\bibfnamefont{T.}~\bibnamefont{Sjostrand}}
  \bibnamefont{et~al.}, \bibinfo{journal}{Comput. Phys. Commun.}
  \textbf{\bibinfo{volume}{135}}, \bibinfo{pages}{238} (\bibinfo{year}{2001}).

\bibitem[{\citenamefont{Adare et~al.}(2009)}]{PhysRevD.79.012003}
\bibinfo{author}{\bibfnamefont{A.}~\bibnamefont{Adare}} \bibnamefont{et~al.}
  (\bibinfo{collaboration}{PHENIX Collaboration}), \bibinfo{journal}{Phys. Rev.
  D} \textbf{\bibinfo{volume}{79}}, \bibinfo{pages}{012003}
  (\bibinfo{year}{2009}).

\bibitem[{\citenamefont{Adler et~al.}(2004)}]{PhysRevLett.92.051802}
\bibinfo{author}{\bibfnamefont{S.~S.} \bibnamefont{Adler}} \bibnamefont{et~al.}
  (\bibinfo{collaboration}{PHENIX Collaboration}), \bibinfo{journal}{Phys. Rev.
  Lett.} \textbf{\bibinfo{volume}{92}}, \bibinfo{pages}{051802}
  (\bibinfo{year}{2004}).

\bibitem[{\citenamefont{Hagedorn}(1965)}]{NuovoCim.Suppl.3.147}
\bibinfo{author}{\bibfnamefont{R.}~\bibnamefont{Hagedorn}},
  \bibinfo{journal}{Nuovo Cim. Suppl.} \textbf{\bibinfo{volume}{3}},
  \bibinfo{pages}{147} (\bibinfo{year}{1965}).

\bibitem[{\citenamefont{Kaplan et~al.}(1978)}]{Phys.Rev.Lett.40.435}
\bibinfo{author}{\bibfnamefont{D.~M.} \bibnamefont{Kaplan}}
  \bibnamefont{et~al.}, \bibinfo{journal}{Phys. Rev. Lett.}
  \textbf{\bibinfo{volume}{40}}, \bibinfo{pages}{435} (\bibinfo{year}{1978}).

\bibitem[{\citenamefont{Adler et~al.}(2007)}]{PhysRevD.76.092002}
\bibinfo{author}{\bibfnamefont{S.~S.} \bibnamefont{Adler}} \bibnamefont{et~al.}
  (\bibinfo{collaboration}{PHENIX Collaboration}), \bibinfo{journal}{Phys. Rev.
  D} \textbf{\bibinfo{volume}{76}} (\bibinfo{year}{2007}).

\bibitem[{\citenamefont{Lafferty and Wyatt}(1995)}]{NIMA.355.541.1995}
\bibinfo{author}{\bibfnamefont{G.~D.} \bibnamefont{Lafferty}} \bibnamefont{and}
  \bibinfo{author}{\bibfnamefont{T.~R.} \bibnamefont{Wyatt}},
  \bibinfo{journal}{Nucl. Instrum. Meth. A} \textbf{\bibinfo{volume}{355}},
  \bibinfo{pages}{541} (\bibinfo{year}{1995}).

\bibitem[{\citenamefont{Buttar et~al.}(2004)}]{Acta.Phys.Pol.B.35.433}
\bibinfo{author}{\bibfnamefont{C.}~\bibnamefont{Buttar}} \bibnamefont{et~al.},
  \bibinfo{journal}{Acta Phys. Pol. B} \textbf{\bibinfo{volume}{35}},
  \bibinfo{pages}{433} (\bibinfo{year}{2004}).

\bibitem[{\citenamefont{Skands}(2010)}]{PhysRevD.82.074018}
\bibinfo{author}{\bibfnamefont{P.~Z.} \bibnamefont{Skands}},
  \bibinfo{journal}{Phys. Rev. D} \textbf{\bibinfo{volume}{82}},
  \bibinfo{pages}{074018} (\bibinfo{year}{2010}).

\bibitem[{\citenamefont{Engel}(1996)}]{Phys.Rev.D.54.4244}
\bibinfo{author}{\bibfnamefont{J.~R.} \bibnamefont{Engel}},
  \bibinfo{journal}{Phys. Rev. D} \textbf{\bibinfo{volume}{54}},
  \bibinfo{pages}{4244} (\bibinfo{year}{1996}).

\end{thebibliography}

\end{document}